\documentclass[acmlarge]{acmart}

\usepackage{booktabs} 
\usepackage{enumerate}
\usepackage{booktabs}
\usepackage{makecell}
\usepackage{multirow}
\usepackage{flushend}
\usepackage{latexsym}
\usepackage{multirow}
\usepackage{amsmath}
\usepackage{amsfonts}
\usepackage{algorithm}
\usepackage[noend]{algpseudocode}
\usepackage{graphicx}
\usepackage{color}
\usepackage{algorithm}  
\usepackage{algpseudocode}  
\usepackage{amsmath}

\newtheorem{defn}{Definition}[section]

\acmJournal{TKDD}
\acmArticle{32}
\acmYear{2021}

\setcopyright{usgovmixed}

\acmDOI{0000001.0000001}

\received{December 2019}
\received{September 2020}
\received[accepted]{January 2021}

\usepackage{filecontents}

\begin{document}

\title{Streaming Social Event Detection and Evolution Discovery in Heterogeneous Information Networks}

\author{Hao Peng}
\affiliation{%
  \institution{Beihang University}
  \city{Beijing}
  \country{China}
  }
\email{penghao@act.buaa.edu.cn}
\author{Jianxin Li}
\authornote{This is the corresponding author}
\affiliation{%
  \institution{Beihang University}
  \streetaddress{37 Xueyuan Rd}
  \city{Haidian}
  \state{Beijing}
  \postcode{100191}
  \country{China}
 }
\email{lijx@act.buaa.edu.cn}
\author{Yangqiu Song}
\affiliation{%
  \institution{Hong Kong University of Science and Technology}
  \city{HongKong}
  \country{China}
 }
\email{yqsong@cse.ust.hk}
\author{Renyu Yang}
\affiliation{%
  \institution{University of Leeds}
  \city{Leeds}
  \country{UK}
  }
\email{r.yang1@leeds.ac.uk}
\author{Rajiv Ranjan}
\affiliation{%
  \institution{Newcastle University}
  \city{Newcastle upon Tyne}
  \country{UK}
 }
\email{raj.ranjan@newcastle.ac.uk}
\author{Philip S. Yu}
\affiliation{%
  \institution{University of Illinois at Chicago}
  \city{Chicago}
  \country{USA}
 }
\email{psyu@uic.edu}
\author{Lifang He}
\affiliation{%
  \institution{Lehigh University}
  \city{Bethlehem}
  \country{USA}
 }
\email{lih319@lehigh.edu}

\begin{abstract}
Events are happening in real-world and real-time, which can be planned and organized for occasions, such as social gatherings, festival celebrations, influential meetings or sports activities.
Social media platforms generate a lot of real-time text information regarding public events with different topics.
However, mining social events is challenging because events typically exhibit heterogeneous texture and metadata are often ambiguous.
In this paper, we first design a novel event-based meta-schema to characterize the semantic relatedness of social events and then build an event-based heterogeneous information network (HIN) integrating information from external knowledge base.
Second, we propose a novel Pairwise Popularity Graph Convolutional Network, named as PP-GCN, based on weighted meta-path instance similarity and textual semantic representation as inputs, to perform fine-grained social event categorization and learn the optimal weights of meta-paths in different tasks.
Third, we propose a streaming social event detection and evolution discovery framework for HINs based on meta-path similarity search, historical information about meta-paths, and heterogeneous DBSCAN clustering method.
Comprehensive experiments on real-world streaming social text data are conducted to compare various social event detection and evolution discovery algorithms.
Experimental results demonstrate that our proposed framework outperforms other alternative social event detection and evolution discovery techniques.
\end{abstract}


\authorsaddresses{
A preliminary version~\cite{peng2019event} of this article appeared in the Proceedings of the 28th International Joint Conference on Artificial Intelligence, Pages 3238-3245 (IJCAI'19).
Authors' addresses: 
Hao Peng, Beijing Advanced Innovation Center for Big Data and Brain Computing, Beihang University, Beijing, China, and with School of Cyber Science and Technology, Beihang University, Beijing, China, \path{penghao@act.buaa.edu.cn}; 
Jianxin Li, Beijing Advanced Innovation Center for Big Data and Brain Computing, Beihang University, Beijing, China, and with State Key Laboratory of Software Development Environment, Beihang University, Beijing, China, \path{lijx@act.buaa.edu.cn};
Yangqiu Song, Department of Computer Science and Engineering, Hong Kong University of Science and Technology, HongKong, China, \path{yqsong@cse.ust.hk};
Renyu Yang, School of Computing, University of Leeds, Leeds, UK, \path{r.yang1@leeds.ac.uk};
Rajiv Ranjan, Computing Science and Internet of Things, Newcastle University, Newcastle, UK, \path{raj.ranjan@newcastle.ac.uk};
Philip S. Yu, Department of Computer Science, University of Illinois at Chicago, Chicago, USA, \path{psyu@uic.edu};
Lifang He, Department of Computer Science and Engineering, Lehigh University, Bethlehem, USA. \path{lih319@lehigh.edu}.
}

\maketitle

\renewcommand{\shortauthors}{H. Peng et al.}

\section{Introduction}\label{sec:introduction}
Nowadays, various social media platforms, such as Weibo, Twitter, Facebook, Instagram, Forum site and Blog, etc., have become major sources for publicizing social events.
With a large amount of events being announced on social media, a large number of comments, reposts and discussions with opinions and emotions from social network users have been generated, and such content can reflect public opinion about many political, economic, security, employment and social welfare, and education issues, etc.
Mining of social media posts, such as online social event detection and evolution discovery, will benefit a lot of real applications, such as predictive analysis~\cite{Radinsky:2013,Asur:2010:PFS,sharmeen2015predicting,liu2019embedding}, disaster risk management~\cite{o2010approaching}, public opinion analysis~\cite{ma2014superedgerank,liu2020event}, information organization~\cite{Allan:2012:TDT:2481012}, recommended systems~\cite{macedo2015context} and others~\cite{blossfeld2014event}.
In general, real-time social event detection and evolution focus on learning high-precision models for identifying event-related clusters from large-scale social messages, as well as fast streaming processing from online social data.

The tasks of real-time social event detection and evolution discovery are more challenging than traditional text mining or social network mining, since a social event is a combination of social user network and text streams in terms of short messages over it. 
More specifically, on the one hand, social events are described in short messages and usually contain keywords and different types of entities, such as person, location, organization, score, date and time, etc~\cite{aggarwal2012event,atefeh2015survey,ji2008refining}. 
This leads to social events being very complicated and heterogeneous.
Besides, events are posted, commented or retweeted by social media users at a specific time.
Thus, modeling social events needs to consider heterogeneous elements as well as social users within social posts.
On the other hand, with the spread of social practices including the participation of new social users and external interventions, events often change over time and streaming nature makes capturing useful semantic information difficult ~\cite{batal2012mining,hienert2012automatic,hong2016building}.
In addition, the variety of events is very diverse and covers a wide range, but the number of events per category is small. 
This feature is very important in social event analysis but has not yet been studied in much detail.
Therefore, due to the above mentioned facts, it is desirable to develop streaming social event detection and evolution discovery methods that consider heterogeneous and dynamic features of events, as well as small sample size problems in each category as a whole.

In the literature, several studies have been aimed at streaming social event detection and evolution discovery. 
However, to our knowledge, none of them consider these two problems in a systematic manner.
The majority of current studies focus on social event detection~\cite{ritter2012open,chandola2009anomaly,becker2011identification,cao2021knowledgepreserving}, leveraging either homogeneous or heterogeneous network.
The first line of thought~\cite{chandola2009anomaly,aggarwal2012event,angel2012dense} is to model streaming social messages as homogeneous words/elements co-occurrence network, and then extract abnormal subgraphs as the events.
Despite the promising results from these studies, their detection accuracies remain unsatisfactory for building reliable and open domain event detection systems in practice.
The second line of thought~\cite{kim2009overview,ritter2012open,ji2008refining,li2013joint} is to model social events as heterogeneous network, which use manually defined frames-based event definitions for extracting social event frames from corpus. 
The frame-based methods can extract entities and their relationships, but can only be used in a limited number of event types, such as earthquake disaster, stock market, venues, politics, etc.
Moreover, most of them are pipeline models consisting of multiple machine learning models to incorporate different levels of annotation and features, and thus their capacities are limited by the components.
For social event evolution discovery, traditional methods mainly focus on exploring evolutionary relationships between events based on various similarity measures, such as TF/IDF~\cite{allan1998line}, KeyGraph~\cite{ohsawa1998keygraph,mori2006topic,sayyadi2009event}, event element based schemes~\cite{lu2015discovering,osborne2014real} and bipartite graph matching methods~\cite{long2011towards}, which allow to infer the strong association relationships between events.
However, similar to the existing detection methods, the above evolution discovery methods are unable to capture the different influences of heterogeneous event elements.

In this paper, we first represent social messages as hyper-edge in an HIN, where all the keywords, entities, topics, social users and time can be connected by this hyper-edge, and define an event meta-schema to characterize the semantic relatedness of social events and build event-based streaming HIN.
In order to enrich the HIN, we extract some information as a complement of the relationships based on the external knowledge base and algorithms.
For accurate social event similarity calculation, we define a weighted \underline{K}nowledgeable meta-paths \underline{I}nstances based \underline{E}vent \underline{S}imilarity measure, namely \emph{KIES}, from semantically meaningful meta-paths.
Second, we then design a novel \underline{P}airwise \underline{P}opularity \underline{G}raph \underline{C}onvolutional \underline{N}etwork model, namely PP-GCN, to learn not only the representation of each event instance, but also optimal weights of meta-paths in event classification and evolution classification tasks, respectively.
Third, in streaming scenario, we use meta-path similarity strategy to extract historical social messages to build a provisional event-based HIN, and then we make use of both the optimal weights of meta-paths, similarity measure \emph{KIES} and a heterogeneous DBSCAN clustering method to detect the social events.
We also estimate the best time delay which is compatible with both the capacity of data acquisition and time consumption of event detection in batch manner, according to experimenting different scales of social messages.
In addition, we also employ both the heterogeneous DBSCAN, namely H-DBSCAN, the optimal weights of meta-paths and event similarity to gather social events with evolutionary relationship.
Finally, in order to speed up the streaming processing of methods, we employ multi-threaded parallel processing technology in the construction of HINs, message searching and clustering to achieve real-time event detection and evolution discovery.
The source code of this work is publicly available at \emph{https://github.com/RingBDStack/social-events}.
The streaming social event detection and event evolution system is online at \emph{http://ring.act.buaa.edu.cn}.

A preliminary version of this work appeared in the proceedings of IJCAI 2019~\cite{peng2019event}.
This journal version has extended the static fine-grained social event categorization model to streaming social event detection and evolution discovery models, and this full version involves several improvements in upgrading methodology and the frame structure of the proposed approach.
We added time element and temporal nearest neighbor relationship when modeling event-based HIN. We also supplemented more explicit explanations of the meta-path in the two scenarios of event detection and event evolution. We presented novel streaming social event detection and evolution framework, and adopted the DBSCAN method - differing from the K-means method that requires to specify the number of clusters - to realize open domain event detection and evolutionary discovery based on density clustering. 
More in-depth experimental results and further analysis are discussed to demonstrate the effectiveness and efficiency of the proposed framework. We supplied the variances of the results of social event mining. For streaming social event mining scenarios, the results on all of the online time consumption and accuracy are displayed. Thorough and deeper analyses are presented, such as effectiveness, efficiency and case study.
In particular, the contributions of this paper are summarized as follows:
\begin{itemize}
\item By modeling social events based on the streaming event-based HIN, the proposed framework integrates event elements including keywords, topic, entities, time, social users and their relations, in a semantically meaningful way, and can also calculate the similarity between any two events.
\item By modeling pairwise popularity graph convolutional network, the model achieves state-of-the-art results in static fine-grained event classification and evolution classification, and learns the optimal weights of meta-paths in different tasks.
\item By learning the optimal weights of meta-paths on a small sample with PP-GCN model, the proposed social event similarity measure $KIES$ based DBSCAN, referred to as H-DBSCAN, achieves new state-of-the-art results for static event detection and evolution discovery tasks, respectively.
\item The combination of multiple meta-paths based social event searching, event similarity and the H-DBSCAN clustering in batch manner achieves state-of-the-art results in streaming social event detection and evolution discovery tasks, respectively.
\item We conduct extensive experiments to demonstrate that the proposed models considerably outperform the state-of-the-art social event detection and evolution discovery methods.
We further illustrate the effectiveness and efficiency of the proposed methods.
\end{itemize}

\section{Heterogeneous Event Modeling}
In this section, we define the problem of modeling social events in heterogeneous information network (HIN) and introduce several related concepts, necessary notations and formulas.

\begin{figure*}[h]
\centering
\includegraphics[width=0.65\textwidth]{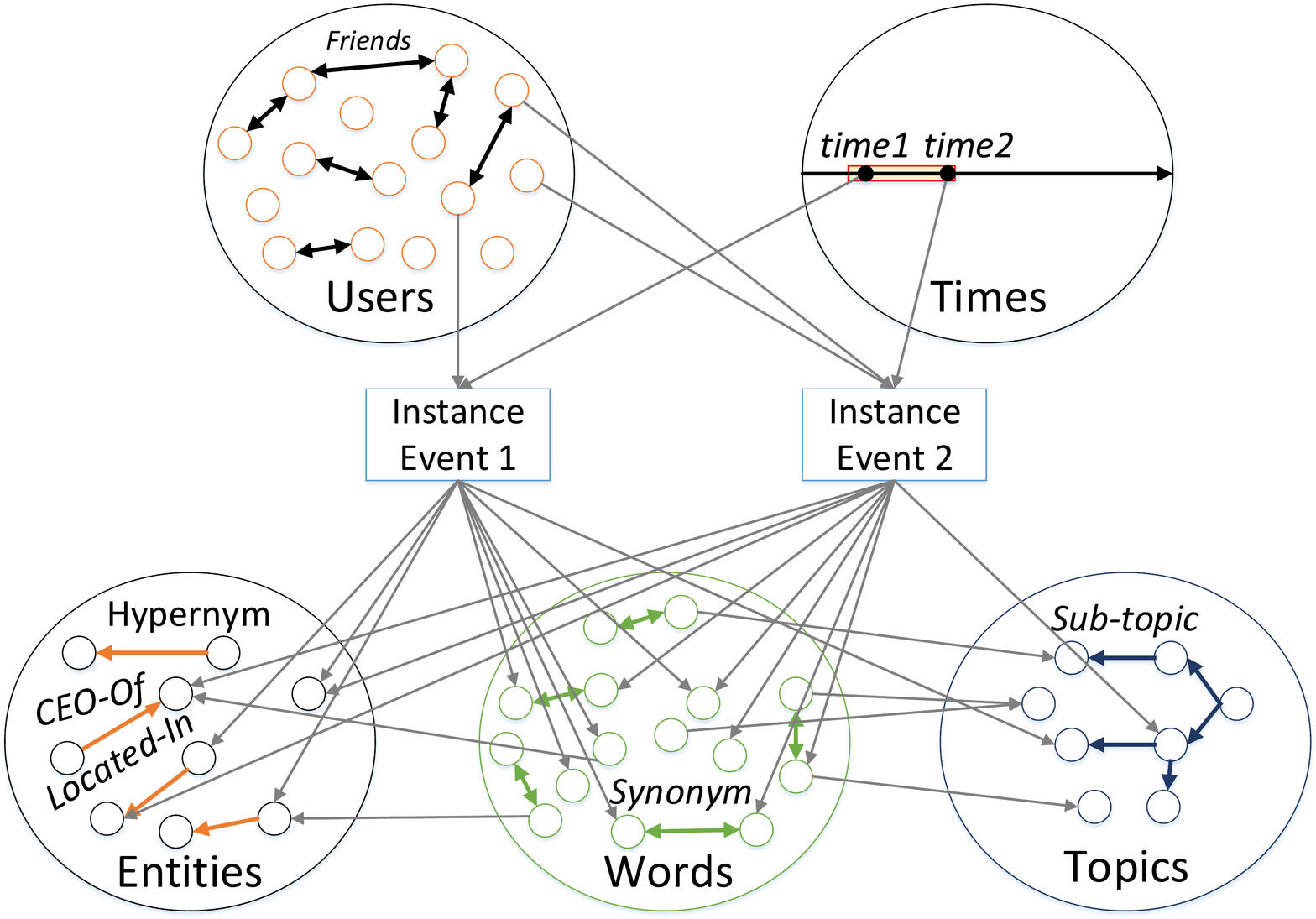}
\caption{Example of two instance events connected by different types of nodes and edges, where instance events are represented as hyper-edges.}\label{fig:event-hin}
\end{figure*}

\begin{figure*}[h]
\centering
\includegraphics[width=0.6\textwidth]{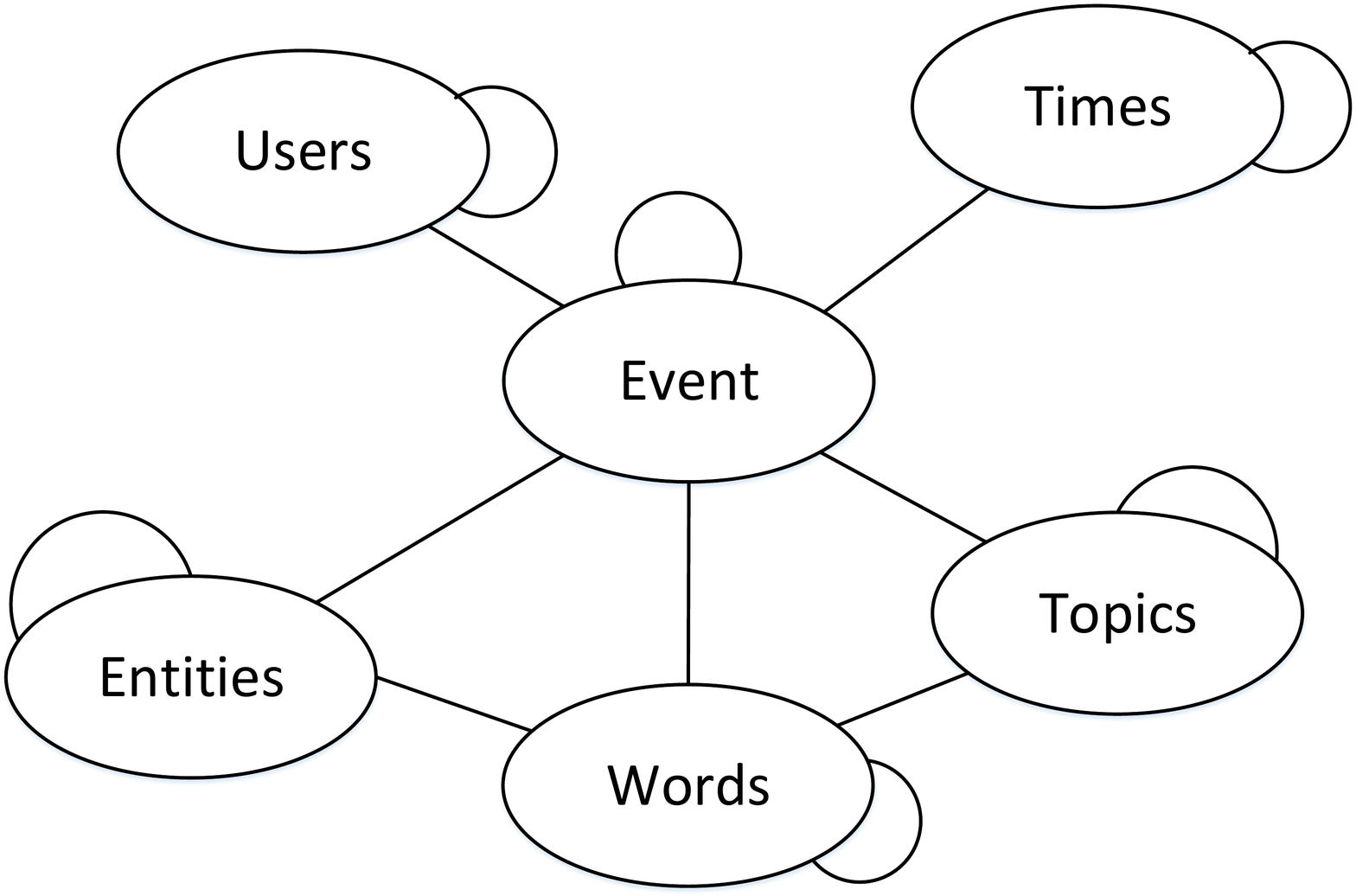}
\caption{Meta-schema of event-based HIN.}\label{fig:event-schema}
\end{figure*}

\subsection{Event Modeling in HIN}
The definition and characterization of ``social event'' have received a great deal of attention in both academic and industrial fields, such as linguistics~\cite{kim2011overview}, cognitive psychology~\cite{zacks2001event}, tourism~\cite{getz2008event}, social networks~\cite{zhou2014event}, etc.
The simplest way to monitor social media events is to represent events as bags-of-words, but it will be more semantically meaningful if we can annotate words and multi-word-expressions as entities with types and different relationships.
Intuitively, a social event generally refers to influential facts that appear on social networks and occur in the real world at some specific time, including creators (posters), named entities such as participants, organizations, festival, specific times, places, currency, address, etc., and other elements such as keywords and topics.
For example, in the tweet ``\emph{Xinhua News Agency, 2019-04-05 19:46: Lightning strike has been confirmed as the cause of a forest fire that killed 27 firefighters and 3 locals in southwest China's Sichuan province, local forest public security bureau said at afternoon of April 5. The provincial forestry and grassland administration cited local police as saying that the lightning igniting the fire was first witnessed and reported by locals in Muli County, Liangshan Yi autonomous prefecture.}'', we can extract the above time, keywords, typed entities and topics from the original social message with NLP tools, and extract social user who post, comment or retweet message from social network.
There are multiple event elements: \textbf{Date}: \emph{April 5, 2019}; \textbf{County}: \emph{Muli}; \textbf{Nation}: \emph{China}; \textbf{Entity 1}: \emph{Lightning Strike} \textbf{Entity 2}: \emph{Local Forest Public Security Bureau}; \textbf{Entity 3}: \emph{Provincial Forestry and Grassland Administration}; \textbf{Entity 4}: \emph{Liangshan Yi Autonomous Prefecture}; \textbf{Entity 5}: \emph{Sichuan Province}; \textbf{Entity 6}: \emph{Southwest China}; \textbf{Topic 1}: \emph{Natural disaster} \textbf{Topic 2}: \emph{Forest fire}; \textbf{Poster}: \emph{Xinhua News Agency}, \textbf{Posting Time}: \emph{2019-04-05 19:46}, etc.
Obviously, the above event's elements are of different types. 
We name the above elements and topics as \emph{event-oriented elements}.
Moreover, in addition to intuitive co-occurrence relationship, after extracting entities, we can make use of external knowledge base~\cite{Auer:2007:DNW:1785162.1785216,xu2017cn,lehmann2015dbpedia} to complement more relationships between entities, such as ``\textbf{located-in}'' relationships among the above locations, ``\textbf{managed-by}'' relationships between the above administrations, etc.
Even within most of the events, there are some relationships between \emph{event-oriented elements}, such as relationships between entities, relationships between keywords, relationships between topics, social relationships between social media users, time relationships, and so on.
We name the above relationships as \emph{event-elements relationships}.
Thus we can model social media events as HIN~\cite{shi2017survey,sun2012mining}.
Next, we introduce how to extract the above rich \emph{event-elements relationships}.

We use the manually organized synonyms to add simple synonym relationships among words in the HIN.
For hierarchical topic structures and the affiliation relationship between keywords and topics in the HIN, we employ the hierarchical latent dirichlet allocation technologies~\cite{griffiths2004hierarchical,blei2010nested} (with about 30 most probable words for each topic).
In order to extract the relationship between entities in the HIN, we consider both accuracy and efficiency, and tackle the problem by following three-steps.
First, we retrieve the entity candidate set with the same name from an external knowledge base, such as the Chinese CN-DBpedia~\cite{xu2017cn} or English Freebase~\cite{lehmann2015dbpedia}.
Second, we use word embeddings~\cite{mikolov2013distributed,mikolov2013efficient} based Word Mover's Distance (WMD) technology~\cite{kusner2015word} to measure the text similarity between the context of entity in the social message and the description of entity in the candidate of the external knowledge base, and then choose the entity with the highest similarity score.
Third, we query the relationship between the above aligned entities from the external knowledge base as the final relationship between entities in the HIN.
In order to establish the relationship between entities and words in the HIN, we extract the keywords in the relevant description of each entity in the external knowledge base and use this affiliation as the relationship between the entity and the keywords.
For the relationship between social users, we consider users with a large number of friends or same interests, and extract friend relationship and shared community relationship~\cite{papadopoulos2012community} between social users in advance.

After extracting the above \emph{event-oriented elements} and \emph{event-elements relationships} from social messages, external knowledge base, resources and tools, we build an event-based HIN, as shown in Figure~\ref{fig:event-hin}, where we name each social message as a instance event.
On the one hand, one social event can usually consist of one unique event instance or multiple strongly associated event instances.
On the other hand, one social event can also be regarded as a co-occurrence of \emph{event-oriented elements}, and the \emph{event-elements relationships} are conducive to explaining the relationship between various elements.
Thus, one social event can be treated as a subgraph of the whole HIN.
To characterize events on social messages, we give an example of the event-based HIN meta-schema, as shown in Figure~\ref{fig:event-schema}.
One particular advantage of the HIN is that meta-paths defined over types (e.g., a typical meta-path ``event-keyword-event'' represents the event similarity based on overlapped keywords between two event instances) can reflect semantically meaningful information about similarities, and thus can naturally provide explainable results for event modeling.

\subsection{Preliminaries}
We introduce some basic definitions based on previous works~\cite{sun2011pathsim,wang2018unsupervised,shi2017survey}, and give some event-HIN related concepts, necessary notations and examples.

\begin{defn}\label{def:hin}
A heterogeneous information network (HIN) is a graph $G = (V,E)$ with an entity type mapping $\phi: V\to A$ and a relation type mapping $\psi: E\to R$, where $V$ denotes the entity set, $E$ denotes the link set, $R$ denotes the relation type set and $A$ denotes the entity type set. 
The number of entity types $|A|>1$ or the number of relation types $|R|>1$.
\end{defn}
For example, Figure~\ref{fig:event-hin} shows an example of two instance events connected with different types of entities, keywords, topics, times, social users and relationships.
After giving a complex HIN for event modeling, it is necessary to provide its meta level (i.e., schema-level) description for better understanding and modeling.

\begin{defn}\label{def:meta-schema}
Given a HIN $G = (V,E)$ with the entity mapping $\phi: V\to A$ and the relation type mapping $\psi: E\to R$, the meta-schema (or network schema) for network $G$, denoted as $T_{G} = (A,R)$, is a graph with nodes as entity types from $A$ and edges as relation types from $R$.
\end{defn}
For example, Figure~\ref{fig:event-schema} shows an example of the HIN meta-schema characterizing events on social messages.
Another important concept is the meta-path which systematically defines relationships between entities at the schema level.

\begin{defn}\label{def:meta-path}
A meta-path $P$ is a path defined on the graph of network schema $T_{G} = (A,R)$ of the form $A_{I}\stackrel{R_{1}}{\longrightarrow}A_{2}\stackrel{R_{2}}{\longrightarrow}A_{3}\cdots A_{L}\stackrel{R_{L}}{\longrightarrow}A_{L+1}$ which defines a composite relation $R=R_{1}\dot R_{2}\dot \cdots \dot R_{L}$ between types $A_{1}$ and $A_{L+1}$, where $\cdot$ denotes relation composition operator, and $L$ is the length of $P$.
\end{defn}
For simplicity, we use object types connected by $\to$ to denote the meta-path when there are no multiple relations between a pair of types: $P = (A_{1}-A_{2}-\cdots-A_{L+1})$.
We say that a meta-path instance $p = (v_{1}-v_{2}-\cdots-v_{L+1})$ between $v_{1}$ and $v_{L+1}$ in network $G$ follows the meta-path $P$, if $\forall l$, $\phi(v_{l}) = A_{l}$ and each edge $e_{l} = <v_{l}, v_{l+1}>$ belongs to each relation type $R_{l}\in P$.
We call these paths as path instances of $P$, denoted as $p\in P$.
$R_{l}^{-1}$ represents the reverse order of relation $R_{l}$.
We will introduce more semantically meaningful meta-paths that describe events in relation to the explanation of meta-paths instances based similarity.

\begin{defn}\label{def:countsim}
Given a meta-path $P = (A_{1}-A_{2}\cdots A_{L+1})$, Cou is a function of the count of meta-path instances such that $Cou_{P}(v_{i},v_{j}) = M_{P}(v_{i},v_{j})$ where $M_{P}=W_{A_{1}A_{2}}\cdot W_{A_{2}A_{3}}\cdots W_{A_{L}A_{L+1}}$ and $W_{A_{k}A_{k+1}}$ is the adjacency matrix between types $A_{k}$ and $A_{k+1}$ in the meta-path $P$.
\end{defn}
For example, the composite relation of two instance events containing the same {event element} and co-occurrence relationship can be described as ``Instance event-Element-Instance event (IEI)'' for simplicity. 
This meta-path simply gives us $M_{IEI} = W_{IE}W_{EI}^{T}$, which is the dot product between event instances, where $W_{EI}$ is the event Instance-Element co-occurrence matrix. 
We can give more event related meta-paths over different lengths, e.g.,
$P_1$: \emph{Event instance-(containing)-Forest Fire-(contained by)-Event instance},
$P_2$: \emph{Event instance-(containing)-Custom Line-(produced by)-Ferretti Group-(contained by)-Event instance},
$P_3$: \emph{Event instance-(containing)-Stephen Curry-(belonging to)-Golden State Warriors-(located at)- Chase Center-(contained by)-Event instance}, etc.
$P_1$ means two event instances are similar if they are containing the same named entity 'Forest Fire'.
$P_2$ means two event instances are similar if they mention named entities 'Custom Line' and 'Ferretti Group', respectively, where Custom Line is a very famous luxury yacht brand produced by the Ferretti Group.
$P_3$ means two event instances are similar if they can be associated by a chain of three event elements with meaningful relationships.
For instance, two event instances contain 'Stephen Curry' and 'Chase Center', respectively, where the home of the Golden State Warriors, to which Stephen Curry belongs, is located at the Chase Center.
Note that the meta-path does not need to satisfy symmetry.
Finally, we can enumerate $22$ symmetric meta-paths in the meta-schema of event-based HIN shown in Figure~\ref{fig:event-schema}.
Here, because a large number of social messages can be generated in a short period of time, we ignore the similarities that occur at the same time.
But, we will implement event detection within the scope of the event instances that occur within a given time period.

Essentially, one social event consists of one or more instance events.
However, in addition to the \emph{event-elements relationships} within the event, there may be evolutionary relationships between events, which can help to reveal more complex and interpretable relationships between events~\cite{makkonen2003investigations,ahirrao2013overview}.
Therefore, the co-occurrence relationship of event evolution can be described as ``Event-Instance event-eleMent-Instance event-Event  (EIMIE)'', where the relation between event and instance event is a composition relationship.
Similar to various meta-paths between instance event, we also present three types of meta-paths that describe the evolution of events over different lengths, e.g.,
$P_1$: \emph{Event-(consisting of)-Instance event-(containing)-Break Prisoner-(contained by)- Instance event-(consisted of)-Event},
$P_2$: \emph{Event-(consisting of)-Instance event-(containing)-Break Prisoner-(synonym of)-Manhunt-(contained by)-Instance event-(consisted of)-Event},
$P_3$: \emph{Event-(consisting of)-Instance event-(containing)-Break Prisoner-(synonym of)-Manhunt-(managed by)- Department of Justice -Instance event-(consisted of)-Event}, etc.
Finally, we also can enumerate $22$ symmetric meta-paths in the meta-schema of event-based HIN shown in Figure~\ref{fig:event-schema}.
Next, we introduce how to normalize the different impacts and the counts of different meta-paths on event similarity.

Similar to the weighted meta-path instances based text similarity~\cite{wang2018unsupervised}, we also define our Knowledgeable meta-paths Instances based social (instance) Event Similarity measure, namely \emph{KIES}.
Intuitively, if two instance events or events are more strongly connected by some important (i.e., highly weighted and meaningful) meta-paths, they tend to be more similar.
Formally, we have
\begin{defn}\label{def:kies}
KIES: a knowledgeable meta-paths instances based social event instance or event similarity. Given a collection of meaningful meta-paths, denoted as $P=\{P_{m}\}_{m=1}^{M'}$, the KIES between two instance events or events $e_i$ and $e_j$ is defined as:
\begin{equation}\label{eq:event_sim_inst}
KIES(e_i,e_j) = \sum_{m=1}^{M'}\omega_{m}\frac{2\times Cou_{P_m}(e_{i},e_{j})}{Cou_{P_m}(e_{i},e_{i})+Cou_{P_m}(e_{j},e_{j})},
\end{equation}
\end{defn}
where $Cou_{P_m}(e_{i},e_{j})$ is a count of meta-path $P_m$ between two instance events or events $e_i$ and $e_j$, $Cou_{P_m}(e_{i},e_{i})$ is that between $e_i$ and $e_i$, and $Cou_{P_m}(e_{j},e_{j})$ is that between $e_j$ and $e_j$.
We use a parameter vector $\vec{\omega} = [\omega_{1},\omega_{2},\dots,\omega_{M'}]$ to denote weights of meta-paths, where $\omega_{m}$ refers to the weight of meta-path $P_{m}$.
$KIES(e_i,e_j)$ is defined in two parts including the semantic overlap and the semantic broadness in the numerators.
Here, the semantic overlap is defined by the number of meta-path instances between two instance events or events $e_i$ and $e_j$, and the semantic broadness is defined by the number of total meta-path instances between themselves.
Therefore, we can calculate a weighted \emph{KIES} distance for any two instance events or events.
Next, we introduce how to learn the different weights $\vec{\omega}$ of meta-paths through limited labeled datasets in event classification and event evolution classification tasks, respectively, by a novel pairwise popularity graph convolution network model.

\section{Pairwise Popularity Graph Convolution Network}\label{sec:PP-GCN}
After building the event-based HIN, we can calculate a weighted similarity for any two instance events or two events by the \emph{KIES} for manually annotated social event data, and then construct an $N\times N$ weighted adjacency matrix $A$, where $N$ is the number of instance events or events and $A_{ij} = A_{ji} = KIES(e_i,e_j)$.
Then we use the unsupervised document representation technology~\cite{le2014distributed} to learn a generalized feature for each event instance or event.
So, we can also construct an $N\times d$ feature matrix $X$, where $d$ is the dimension of the generalized feature.
It is quite clear that, so far, we can make use of the popular GCN architectures~\cite{kipf2017semi,defferrard2016convolutional} to learn discriminating event representation and weights of meta-paths based on both the interactions among events and generalized event features in node classification task.
Here, the input of the GCN model includes both $A$ and $X$ matrices, and one class represents one social event class or one social event evolution class.

In order to construct a preliminary GCN model, we utilize the popular multi-layer GCN architecture~\cite{kipf2017semi} with the following layer-wise propagation rule: 
\begin{equation}\label{eq:obj_gcn}
H^{(l+1)} = \sigma(\widetilde{D}^{-\frac{1}{2}}\widetilde{A}\widetilde{D}^{-\frac{1}{2}}H^{(l)}W^{(l)}),
\end{equation}
where $\widetilde{A} = A + I_{N}$, $\widetilde{D}$ is diagonal matrix such that $\widetilde{D}_{ii}=\sum_{j}\widetilde{A}_{ij}$, $I_{N}$ is the identity matrix, $W$ is the parameter matrix, and $l$ is the number of layers.
Next, let $Z$ be an output $N\times F$ feature matrix, where $F$ to be the dimension of output representation per event instance.
The input layer to the GCN is $H^{(0)} = X, X\in R^{N\times d}$, which contains original event instance feature, $H^{(l)} =Z$, and $Z$ is graph-level output.
And $\sigma$ denotes an activation function such as Sigmoid or ReLU.
Note that we also divide the $N$ event instances or events into two parts: the training set and the testing set.

\begin{figure*}[t]
\centering
\includegraphics[width=0.98\textwidth]{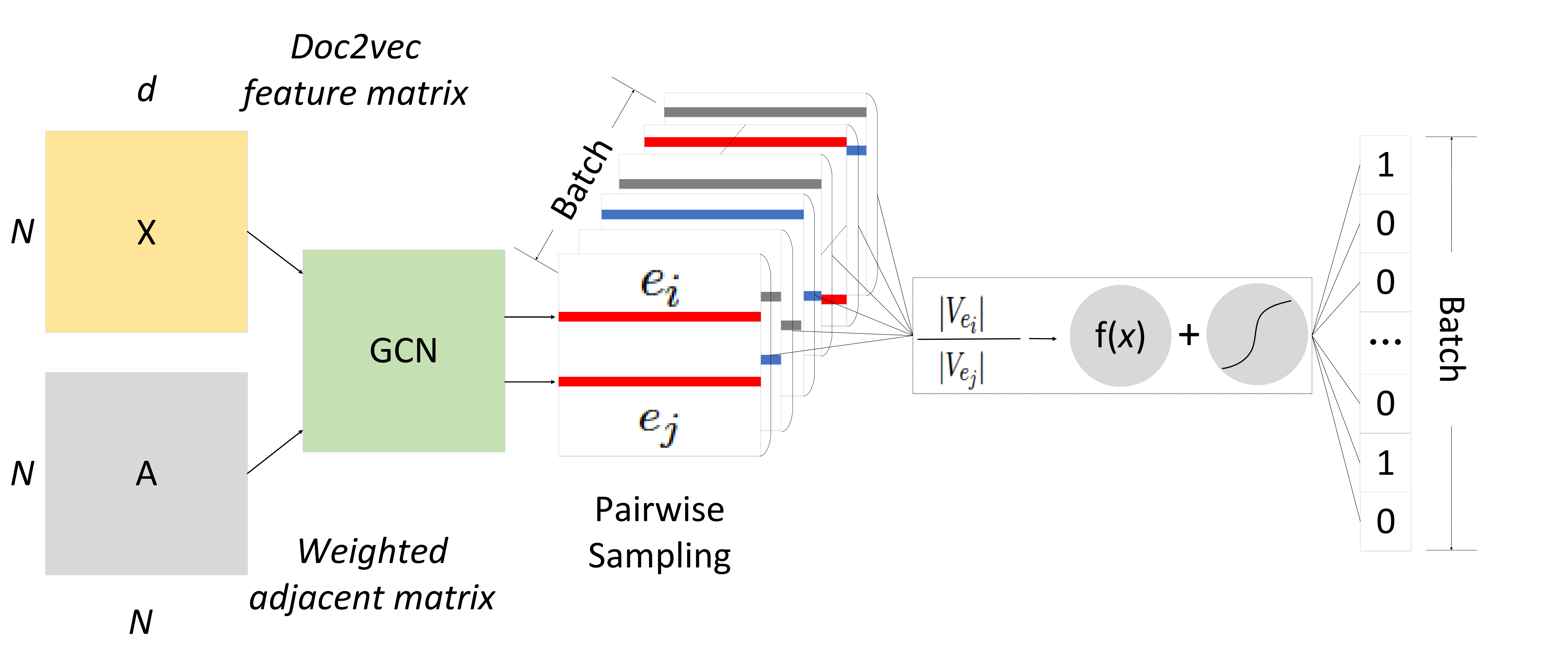}
\caption{An overview of the proposed Pairwise Popularity Graph Convolutional Network (PP-GCN).}\label{fig:pp-gcn}
\end{figure*}

However, the real-world and real-time social events naturally have two problems of sparsity: the small number of  instance events for each event and a large number of categories.
So, we sample instance events pair, and judge whether the pair belongs to one event to train instance event representation by a pairwise GCN model.
For event evolution, we also sample the event pair, and judge whether the pair belongs to one event evolution to train event representation by the pairwise GCN model.
As shown in Figure~\ref{fig:pp-gcn}, we present the proposed pairwise GCN model.
Before explaining the pairwise GCN model, we show how to implement a pairwise sampling to generate training samples.
We assume that if a pair of instance events $e_i$ and $e_j$ from a training set belong to the same class, we name the pair $e_i$ and $e_j$ as a \emph{positive-pair} sample.
If a pair of instance events $e_i$ and $e_j$ from training set belong to two different classes, we name the pair $e_i$ and $e_j$ as a \emph{negative-pair} sample.
As shown in Figure~\ref{fig:pp-gcn}, if the pair is a {positive-pair} sample, we represent its by two red lines; if the pair is {negative-pair} sample, we use both gray line and blue line to represent it.

After explaining how to generate training samples, we take instance event classification as an example to introduce the pairwise GCN model.
Firstly, we randomly select $R$ (i.e., 1024) instance events from training set of the original $N$ samples as a preliminary set, then randomly select two instance events for each instance event in the set to construct one positive-pair sample and one negative-pair sample, and finally we construct a $2R$ instance event pairs set to train the pairwise GCN model.
Here, both the {positive-pair} and {negative-pair} samples are equal to $R$.
Second, we randomly sample $B$ (i.e., 64) samples from the $2R$ (i.e., 2000) instance event pair set to construct one batch to forward propagation of our proposed model, as shown in Figure~\ref{fig:pp-gcn}.
Third, the second step is cycled $E$ (i.e., 64) times to form an epoch.
For the next epoch, we loop through the above three steps to train the GCN model and learn discriminating instance event representation and weights of meta-paths with instance event classification tasks.
However, the above pairwise sampling based GCN model can not guarantee that the model has a smooth convergence and avoids overfitting during training.
In order to avoid the uncertainty fluctuation of learning weights, we introduce a novel pairwise popularity GCN model to train it.

Taking instance event classification as an example, we assume that any event class has an average of $r$ event instances.
The probability that any event instance selected as a {positive-pair} sample is about $\frac{1}{r}$, and the probability of being selected as a {negative-pair} sample is about $\frac{1}{N-r}$.
We note that $\frac{1}{r} \gg \frac{1}{N-r}$ in general.
It is obvious that the {negative-pair} samples have more diversity than the {positive-pair} samples.
Most previous works~\cite{papadopoulos2012popularity,weng2013role,cannistraci2013link} have observed the phenomenon that the connected probability of a sample determines the popularity of it.
The greater the probability of the connection, the greater the popularity of the features of the sample.
Inspired by this observation, we assume that in feature representation learning, the modulus of the learned feature vector is also larger if the popularity is greater.
So, the two modulus of learned event instance feature vectors of {positive-pair} will be closer.
Similarly, the angle between the features of the two samples represents the semantic distance, and the smaller the angle, the closer the semantics.
However, the range of angle between any two event instance features is $[0, \frac{\pi}{2}]$, but the range of the ratio of any two modulus can be $[1, +\infty)$.
We believe that the ratio of modulus based pairwise popularity GCN model can generate more discriminating instance event representation than the angle based pairwise GCN, namely PA-GCN, as validated in the experiment section.
Next, we introduce how to design a suitable neural network layer of the output, and calculate the loss function for the GCN model.

For more discriminating feature representation learning of our GCN model, we make use of the popularity of the output instance event feature vectors in $Z$ to distinguish different classes.
As shown in the Figure~\ref{fig:pp-gcn}, for any two learned instance event vectors $V_{e_{i}}$ and $V_{e_{j}}$ that satisfy $|V_{e_{i}}|\ge |V_{e_{j}}|$, we employ a ratio of modulus $x=\frac{|V_{e_{i}}|}{|V_{e_{j}}|}$ as the input of a nonlinear mapping function $f(x) = - log(x-1+c)$, where the coefficient $c$ is 0.01 to limit the upper bound of output of the function.
Note that we assume that the ratio of modulus of \emph{positive-pair} belongs to $[1, 2)$, and the ratio of modulus of \emph{negative-pair} belongs to $[2,+\infty)$.
So, the nonlinear mapping function $f(x)$ can map the above ratio $x$ from [1, 2) to (0, 2], and [2, +$\infty$) to (-$\infty$, 0).
Next, we add a Sigmoid function to map the output of the nonlinear mapping layer to 0 or 1 by a threshold 0.5.
As shown in the Figure~\ref{fig:pp-gcn}, one \emph{positive-pair} or \emph{negative-pair} input sample can only be paired with an output of 0 or 1.
For one batch (64 pairs) samples, our model can generate one batch size ($1\times 64$) of a one-zero output vector.
It stands to reason that we can use a cross entropy function as our GCN model's loss function, and employ the popular stochastic gradient descent (SGD) method to iterate all parameters to train the model.
The learned weights also will be used to measure similarity for any two social events or instance events, respectively, by different classification tasks of event evolution classification or event classification.
To verify the smooth convergence and avoidance of overfitting ability of our model, we can perform over 7000 epochs, and observe that the evaluation criteria of the model changes over time in Section~\ref{sec:expe}.

To evaluate the effectiveness of the PP-GCN model, we design the following steps to test any instance event or event sample $t$ from the test set of the original $N$ samples.
We first assume that there is a total of $C$ event or event evolution classes in the original $N$ instance events.
Secondly, we calculate the ratio of the modulus of the representation vectors for sample $t$ and the remaining $N-1$ samples, respectively.
If the ratio of the modulus falls within the interval of $[1,2)$, we consider the pair belongs to the same class.
Otherwise, if the ratio of the modulus is greater than 2, we consider the pair belongs to different classes.
Then, for each event or event evolution class, we can get a probability that the sample $t$ most likely belongs to it.
Finally, we select the event class with the highest probability as the test output for the sample $t$.
Note the fact that the event category of the test set may not be included in the event category of the training set.

After the previous analysis, we can calculate the similarity for any two instance events under the event-HIN framework and the weights $\vec{\omega}$.
We also calculate the average similarity threshold $\theta_e$ between the instance events that make up the event, and the average similarity threshold $\theta_s$ between the events that make up the event evolution, from the manually annotated event dataset.
Since our proposed meta-paths and similarity measure \emph{KIES} have better interpretability, we also implement event clustering based on the learned weights of meta-paths and the \emph{KIES} for event detection and event evolution tasks, respectively.
Next, we introduce how to implement streaming event detection and event evolution based on the technologies of meta-paths based social messages searching, event-HIN, the weights $\vec{\omega}$, similarity measure \emph{KIES} and clustering algorithm.

\section{Streaming Social Event Detection and Evolution}\label{sec:detection-evolution}
In this section, we introduce how to implement streaming social event detection and event evolution based on the proposed technologies and multi-core parallel heterogeneous DBSCAN algorithm~\cite{ester1996density,patwary2012new}, respectively.
For real-world original streaming social messages, we divide the daily streaming social messages into time slices evenly according to the acquisition time.
However, the occurrence of a social event is likely to span multiple neighbouring time slices, in other words, a social event may cover social messages in multiple time slices.
Similarly, a social event evolution set often covers more events over longer neighboring time slices.
Thus, we introduce a meta-paths based social messages and event searching method to build provisional event-based HIN.
Third, we propose the multi-core parallel heterogeneous DBSCAN (H-DBSCAN) method based on two distances including the weights learned by PP-GCN's static event classification and event evolution classification tasks, respectively, to achieve event detection and evolution discovery.

\subsection{Meta-paths guided searching on HIN}\label{sec:hin_searching}
\begin{figure*}[t]
\centering
\includegraphics[width=0.9\textwidth]{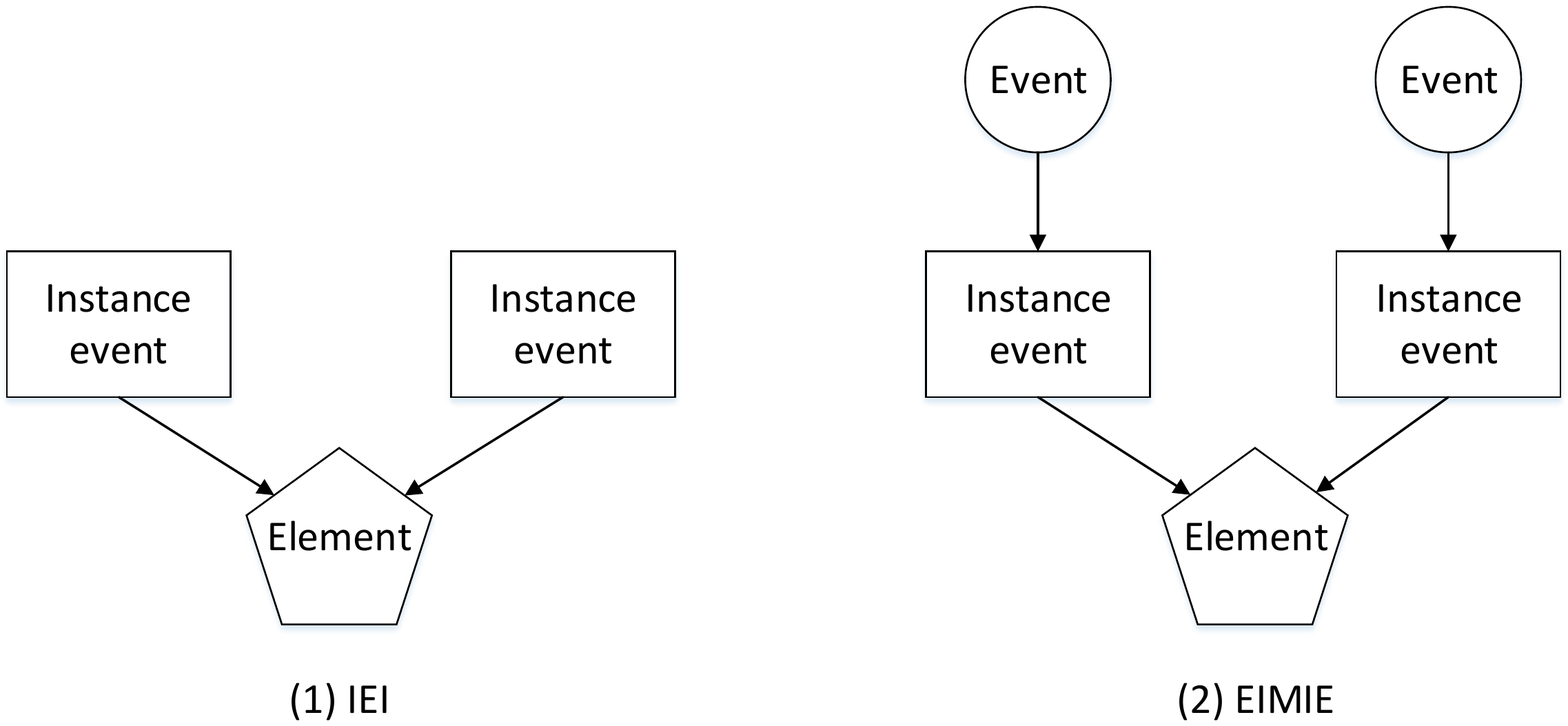}\vspace{-0.85in}
\caption{Examples of two meta-paths in event-based HIN.}\label{fig:metapaths}
\end{figure*}

As shown in Figure~\ref{fig:metapaths}, we present two types of meta-paths that describe the relationship between social instance events, and the relationship between social events.
As discussed in the example of the Definition~\ref{def:countsim}, if any two social messages or social events can be connected by any instances of above meta-paths, we consider them to have interpretable relationships.
In order to reduce the number of historically relevant social messages and consider only the real situation of social events, we limit the time span for searching social data.
In the case of event detection, given any instance event $E_i$ in the current $t$ time slice, we collect the historically relevant social event instance $E_j$ meeting:
\begin{equation}\label{eq:search_messages}
\begin{aligned}
\Delta T = E_{i}.time - E_{j}.time < T_{1}\quad \text{and}\quad  Score = E_{i}.elements \cap  E_{j}.elements \neq \{\varnothing\},
\end{aligned}
\end{equation}
where $T_{1}$ refers to the longest time span for event detection.
We note the current streaming social messages collection that combines latest social messages in the $t$ time slices and historical relevant social messages, as $CM(t)$.
Similar to event evolution, given any detected event $E_m$ in current time slice, when we collect the historically relevant social event $E_n$ meeting:
\begin{equation}\label{eq:search_events}
\begin{aligned}
\Delta T = E_{m}.time - E_{n}.time < T_{2}, \quad T_{1} < T_{2}\quad \text{and}\quad  
Score = E_{m}.elements \cap  E_{n}.elements \neq \{\varnothing\},
\end{aligned}
\end{equation}
where $T_{2}$ refers to the longest time span for event evolution.
We denote the current streaming social events collection that combines latest social events in the $t$ time slice and historical relevant social events, as $CE(t)$.
Since the more relevant the event, the greater the correlation $Score$, we also limit the number of instance events or events retrieved based on the ranking of correlation $Score$.
We have noticed that the computational complexity of the query in Eq.~\ref{eq:search_messages} is $O(n)$ in the HIN.
And, the computational complexity of the query in Eq.~\ref{eq:search_events} is $O(kn)$, where $k$ refers to the average number of instance events contained in an event.
Next, we present how to implement streaming event detection and evolution discovery based on $CM(t)$ and $CE(t)$, respectively.

\subsection{Parallel H-DBSCAN based streaming event detection and evolution discovery and system}\label{sec:dbscan}
Social events are regarded as a co-occurrence of event elements including themes, dates, locations, people, organizations, keywords, social behavior participants and time, etc.
And social events are the unique aggregation of a variety of semantics.
We believe that the same event or the evolution tends to be cohesive, and there is a discernible distance between the semantics of different events or evolutions.
So, we propose heterogeneous DBSCAN (H-DBSCAN), which is the \emph{KIES} distance based DBSCAN.
The main advantages of H-DBSCAN include that 
1), it does not require one to specify the number of clusters in the data a priori; 
2), it can find arbitrarily shaped clusters, and can even find a cluster completely surrounded by a different cluster;
and 3), it has a notion of noise, and is robust to outliers.
Intuitively, the H-DBSCAN method is very suitable for event clustering.

\begin{figure*}[t]
\centering
\includegraphics[width=0.9\textwidth]{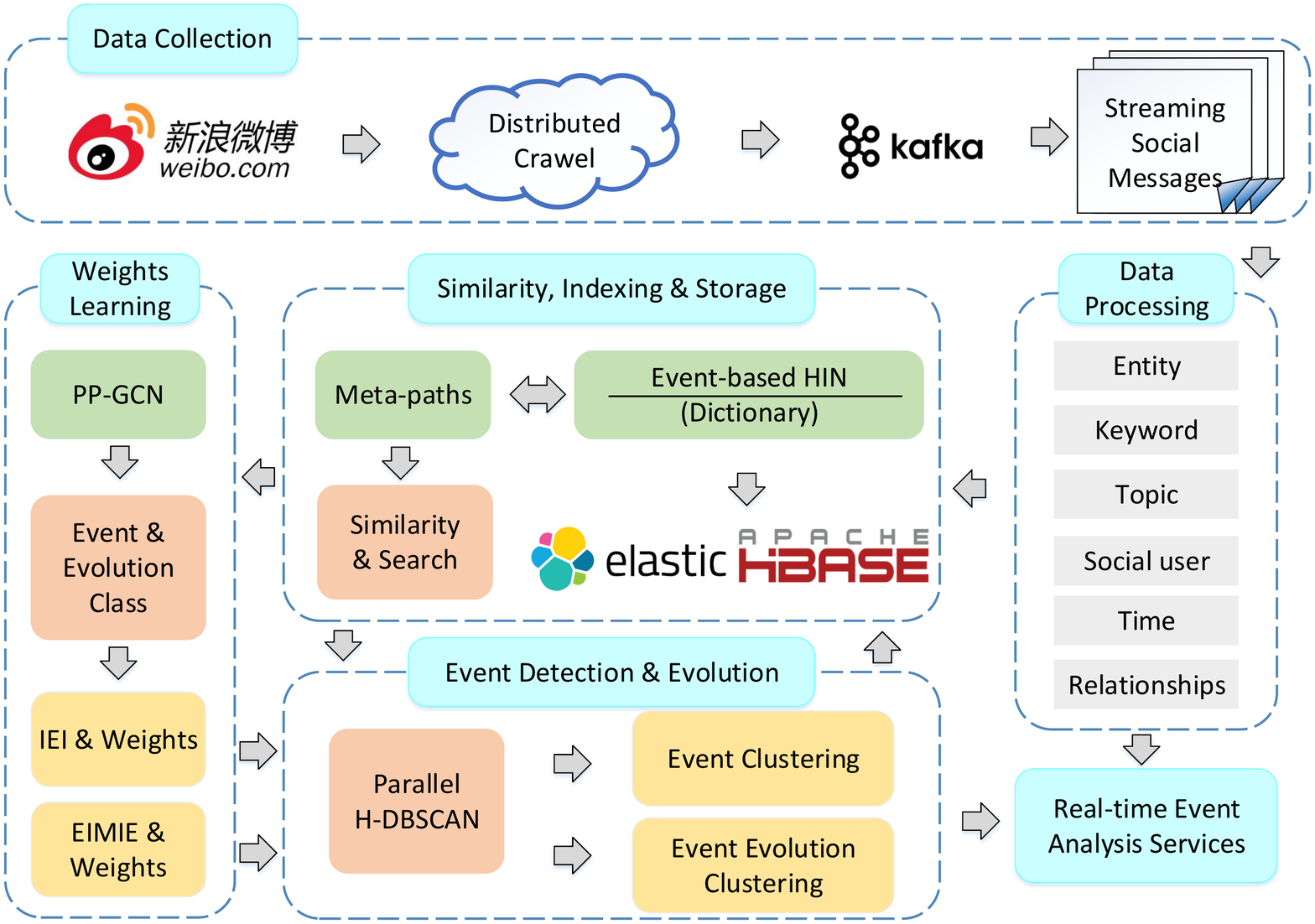}
\caption{Architecture of streaming social event detection and evolution discovery.}\label{fig:architecture}
\end{figure*}

At the $t$ time slice, the input of H-DBSCAN module includes the current streaming social messages: $CM(t)$ or $CE(t)$, distance threshold within a cluster: $\epsilon$ and the minimum number of points required to form a cluster: $minPts$.
Compared to mapping the social data into coordinate space, it is more convenient to use a similarity distance-based adjacency matrix to model the near-far relationship among the social messages.
For instance, for any two social messages $e_i$ and $e_j$ in the $CM(t)$, we build a distance-based adjacency matrix $DA(t)$, where $DA(t)_{ij} = DA(t)_{ji} = 1 - KIES(e_i,e_j)$, to represent the social data.
For the given distance threshold $\epsilon$, we choose the best similarity based on the value of the \emph{KIES} distances of the manually labeled same event (evolution) and different events (evolutions).
For the given of the minimum number of points $minPts$, we choose the value of 1, due to the sparsity caused by the streaming data collect, a unique social event instance may be a separate social event. 
For faster and streaming social event detection and evolution discovery, we also use the multi-core parallel H-DBSCAN algorithm using $p$ threads.

The pseudocode of the H-DBSCAN algorithm for social event detection is given in Algorithm~\ref{alg:dbscan}.
Since the distance between samples is already stored in the adjacency matrix $DA(t)$, the worst computational complexity of the function of the query neighbors is $O(n)$.
However, we employ an indexing structure to store the matrix, the computational complexity of the query function is $O(log n)$ in line~\ref{code:dbscan:getneighbors1} and ~\ref{code:dbscan:getneighbors2}.
For evolution discovery, similar to event detection, the input parameters are the adjacency matrix $DA(t)$ based on $CE(t)$, $\epsilon$, and $minPts$ of the event evolution task.
Therefore, the overall average runtime complexity of the parallel H-DBSCAN is $O(\frac{n log n}{p})$.

\begin{algorithm}[htb]  
  \caption{Framework of parallel H-DBSCAN algorithm for events clustering.}\label{alg:dbscan}
  \begin{algorithmic}
    \Require
      The current streaming social data $CM(t)$;
      The \emph{KIES} distance-based adjacency matrix $DA(t)$;
      The \emph{KIES} distance threshold within a cluster $\epsilon$;
      The minimum number of points required to form a cluster $minPts$;
      The number of threads $p$;
    \Ensure
      A set of clusters.
    \For{$x$ = 1 to $p$ in parallel}
        \For{any unvisited event instance $e_i \in CM(t)$}
          \State mark $e_i$ as visited
          \State $nS \gets GetNeighbors(DA(t), e_i, \epsilon$)\label{code:dbscan:getneighbors1}
          \If {$|nS| < minPts$}
            \State mark $e_i$ as noise
          \Else
            \State $C \gets \{e_i\}$
            \For{each event instance $e_j \in nS$}
              \State $nS \gets nS\backslash e_j$
              \If {$e_j$ is not visited}
                \State mark $e_j$ as visited
                \State $nS' \gets GetNeighbors(DA(t), e_j, \epsilon$)\label{code:dbscan:getneighbors2}
                \If {$|nS'| \geq minPts$}
                    \State $nS \gets nS\cup nS'$
                \EndIf 
              \EndIf
              \If {$e_j$ is not yet member of any cluster}
                \State $C \gets C \cup \{e_j\}$
              \EndIf  
            \EndFor
          \EndIf
        \EndFor
    \EndFor 
  \end{algorithmic}
\end{algorithm}

The architecture of streaming social event detection and event evolution is shown in Figure~\ref{fig:architecture}.
This architecture mainly consists of six modules, namely \emph{data collection}, \emph{data processing}, \emph{similarity, indexing $\&$ storage}, \emph{weights learning}, \emph{event detection $\&$ evolution} and \emph{real-time event analysis services}.
For the \emph{data collection}, a distributed crawler is developed to continuously fetch open microblogs and social network information through Weibo webpage~\footnote{https://www.weibo.com}.
The collected data is forwarded to processing modules through Kafka~\footnote{https://kafka.apache.org} to decouple their dependency.
For the \emph{data processing}, we extract both \emph{event-oriented elements} and \emph{event-oriented relationships} to construct streaming event-based HIN, where we store the edges into multiple dictionaries.
For the \emph{similarity, indexing $\&$ storage}, in addition to the streaming event-based HIN, it also provides meta-paths instances based similarities, meta-paths based searching, and HBase~\footnote{https://hbase.apache.org} and Elasticsearch~\footnote{https://www.elastic.co} for data storage and full-text indexing.
For the \emph{weights learning}, a semi-supervised pairwise popularity graph convolutional network is proposed to learn the best weights of meta-paths for event detection and evolution, respectively.
For the \emph{event detection $\&$ evolution}, parallel H-DBSCAN is implemented for event detection and evolution discovery, respectively.
For the \emph{real-time event analysis services}, it presents real-time event detection and related interpretation based on event elements and event relationships, as well as real event evolution representation.

\section{Experimental Evaluation}\label{sec:expe}
In this section, we evaluated both effectiveness and efficiency of our proposed methods and streaming system by several real-world event detection and event evolution tasks over state-of-the-art models.
What's more, we implement case study of real-world event detection and event evolution, combining \emph{event-oriented elements}, \emph{event-oriented relationships} and event evolutionary relationship to give more interpretation and analysis of social events.

\subsection{Datasets, experimental settings and evaluation metrics}
We select popular and open social media platforms Sina Weibo, a hybrid of Twitter and Facebook, and the Twitter of China and Chinese Social Media to collect richer social messages datasets.
Each event instance is a non-repeating social message text.
One event is a set of instance events that contain semantically identical information revolving around a real world incident.
An event always has a specific time of occurrence.
It may involve a group of social users, organizations, participating persons, one or several locations, other types of entities, keywords, topics, etc.
One event evolution is a set of events that are incident in a sequence and describe the specific events occurring at different stages of the event.
Our distributed data collection platform collects an average of $2.35$ million Weibo after filtering noise and removing duplication per day.
A total of 100,000 meaningful Weibo social messages are manually labeled for ground truth, and both event and event evolution labels for the collected Weibo messages are labeled by the outsourcing companies.
For example, social media's tweet about \emph{Tiger Woods winning the 2019 Masters of Golf} is an influential event in the real world and unlike \emph{Patrick Reid's 2018 Masters of Golf}. 
These are two different events that happen in the real world and belong to different event categories.
Note that the social users and their friend relationships involved in Sina datasets are granted by the Sina company for scientific research purposes only.

In order to learn the weight of meta-paths and implement static event and event evolution classifications.
We use 10000 manually labeled Weibo social data, where 60\% of the data as training set, 20\% of data as development set and the remaining 20\% of as test set.
The number of the event classes is 5,470, and the number of event evolution classes is 5,260.
We can see that the total number of classes is large and the number of samples in each class is small.
In order to evaluate the effectiveness and efficiency of the proposed event-based HIN, meta-paths instances based similarity and parallel H-DBSCAN clustering for static social event detection and evolution discovery, we use social data ranging from 10,000 to 100,000.
In order to evaluate the effectiveness and efficiency of the streaming social event detection and evolution discovery, we use the online 4.2 million Weibo data per day.

All of the contrast experiments are performed on three servers cluster, each with 64 core Intel Xeon CPU E5-2680 v4@2.40GHz with 512GB RAM and 4$\times$NVIDIA Tesla P100-PICE GPUs, and an external shared configuration.
The operating system and software platforms are Ubuntu 5.4.0, Tensorflow-gpu (1.4.0) and Python 2.7.
The metrics used to evaluate the effectiveness of event detection and event evolution classifications are the accuracy and F1 score. 
The metric used to evaluate the effectiveness of event detection and evolution discovery is the normalized mutual information (NMI). 
For event detection and event evolution classification tasks, we train more than 7000 epochs. 
We then choose the weights of meth-paths with the best test performances.

\subsection{Compared methods}\label{sec:baselines}
Since it includes three parts of works, we briefly describe the text matching based static social message or social event classification methods, document
similarity distances, and streaming social event detection and event evolution methods.

The following eight comparative models are the static social message or social event classification methods.
\begin{itemize}
    \item \textbf{Support Vector Machine with TF-IDF feature (SVM)}: Support Vector Machine with pair document TF-IDF features is the most classical approach for classification tasks. In this approach, we extract the TF-IDF features for social messages and social events, respectively, and then use the SVM classifier to implement the static event and event evolution classification.
    
    \item \textbf{Convolutional Matching Architecture-I (ARC-I)}~\cite{hu2014convolutional}: It encodes text pairs by convolution neural networks capturing the rich matching patterns at different levels, and compares the encoded representations of each text with a multi-layer perception.

    \item \textbf{Convolutional Matching Architecture-II (ARC-II)}~\cite{hu2014convolutional}: It builds directly on the interaction space between two texts, and models all the possible combinations of them with 1-D and 2-D convolution neural networks with softmax function.

    \item \textbf{Match by Local and Distributed Representations (DUET)}~\cite{mitra2017learning}: It builds text ranking model composed of two separate deep neural networks, one that matches two texts using a local representation, and another that matches the two texts using learned distributed representations. The two networks are jointly trained as part of a single neural network. 

    \item \textbf{Multiple Positional Semantic Matching (MV-LSTM)}~\cite{wan2016deep}: It matches two texts with multiple positional text representations, where each positional sentence representation is a sentence representation at this position generated by a bidirectional long short term memory. The output score is finally produced by aggregating interactions between these different positional sentence representations, through k-Max pooling and a multi-layer perceptron.

    \item \textbf{Convolutional Deep Structured Semantic Models (C-DSSM)}~\cite{shen2014learning}: It builds a latent semantic model based on a convolutional neural network to learn low-dimensional semantic vectors for texts, where convolution-max pooling operation, local contextual information at the word n-gram level and salient local features in a word sequence are used.
    
    \item \textbf{Deep Structured Semantic Model (DSSM)}~\cite{huang2013learning}: It utilizes deep neural networks to map high-dimensional sparse features into low-dimensional features, and calculates the semantic similarity of the document pair. The deep models are discriminatively trained by maximizing the conditional likelihood. 
    
    \item \textbf{Siamese Encoded Graph Convolutional Network (SE-GCN)}~\cite{liu2018matching}: It learns vertex representations through a Siamese neural network and aggregates the vertex features though GCNs to generate the document matching.

\end{itemize}

The following six comparatives are state-of-the-art document similarity distance.

\begin{itemize}

    \item \textbf{Term Frequency-Inverse Document Frequency (TF-IDF)}: It uses the bag-of-words representation divided by each word's document frequency, which reflects how important a word is to a document in a text collection.

    \item \textbf{Latent Dirichlet Allocation (LDA)}~\cite{blei2003latent}: It is a generative statistical model for text documents that learns representations for documents as distributions over word topics, and allows sets of observations to be explained by unobserved groups that explain why some parts of the data are similar.

    \item \textbf{Marginalized Stacked Denoising Autoencoder (mSDA)}~\cite{chenmarginalized}: It consists of multiple stacked layers of denoising autoencoders, and is a representation learned from stacked denoting autoencoders.

    \item \textbf{Componential Counting Grid (CCG)}~\cite{perina2013documents}: It is a generative model that models documents as a mixture of word distributions and LDA,  using the counting grid embedding through window overlapping.

    \item \textbf{Word Move Distance (WMD)}~\cite{kusner2015word}: It measures the dissimilarity between two documents as the minimum amount of distance that words of one document need to travel to reach words of another document.

    \item \textbf{Knowledge-driven document similarity measure (KnowSim)}~\cite{wang2016text}: It's also a meta-paths instances based document similarity, and hasn't considered the impacts of social users. The weights of meta-paths are estimated by the Laplacian scores of documents.

\end{itemize}

The following nine comparative metrics are mainstream social event detection and evolution clustering methods.

\begin{itemize}
    \item \textbf{Hashtag-based Peak Aggregation based Event Detection (HPA-ED)}~\cite{Marcus2011Twitinfo}: It detects events by aggregating volume peaks of hashtag over time in social streams.

    \item \textbf{Streaming Keyword Co-occurrence Graph based Event Detection and Event Evolution (SKCG)}~\cite{sayyadi2009event,sayyadi2013graph}: It builds streaming keywords co-occurrence network, and employs a betweenness centrality based clustering algorithm to detect social events. 

    \item \textbf{Combination of Social Content and online social Behavior based Event Detection (CSCB-ED)}~\cite{Nguyen2017Real}: It combines content-based features and the propagation of news to detect social events. We implement it by treating forwarding networks as friendship networks.

    \item \textbf{Streaming and Bursting Abnormal Subgraph based Event Detection (SBAS-ED)}~\cite{yu2017ring}: It combines trending keywords, overlapping community detection and streaming distributed processing to discover social events.

    \item \textbf{Inverted indices and Incremental Clustering based Event Detection (IIC-ED)}~\cite{hasan2019real}: It incorporates specialized inverted indices and an incremental clustering approach to provide a low computational cost solution to detect both major and minor newsworthy events in real-time from the social data stream.

    \item \textbf{Content Similarity and Temporal Proximity based Event Evolution (CSTP-EE)}~~\cite{nallapati2004event}: It combines content similarity and temporal proximity to measure the relationships between
events. 

    \item \textbf{Sketch Graphs Tracking based Event Detection and Event Evolution (eTrack)}~\cite{Lee2013Event}: It models the social streams as an evolving network, where each social post is a node, and edges between posts are constructed when the post similarity is above a threshold, and tracks events by extracting (k,d)-core subgraph.
The sketch graph is subgraph induced by social messages' core posts and core edges.

    \item \textbf{Sliding Inverse Document Frequency of terms based Event Detection and Event Evolution (SIDF)}~\cite{weiler2014event}: It models a simple event identification approach, which uses a sliding window model to extract events and the context of events in real-time from social message texts. This approach is based on monitoring shifts in the inverse document frequency (IDF) of terms.

    \item \textbf{Weighted and Locally Sensitive Hash based Event Evolution (WLSH-EE)}~~\cite{lu2015discovering}: It integrates both linear weighting of event-elements and locally sensitive hash based distance metric to cluster the event evolution chain.
\end{itemize}

\subsection{Effective evaluation of weights learning}~\label{sec:weightslearning}
In order to calculate the similarity between any two social instance events or events with the meth-paths based distance measure in Eq.~\ref{eq:event_sim_inst}, we design the PP-GCN model to learn the different weights and perform static social event classification and event evolution classification, respectively.
Table~\ref{tab:event-classification} shows the testing accuracy and F1-score of different algorithms on the tasks of event classification and event evolution classification in the 2000 testing Weibo data.
Overall, the proposed PP-GCN model consistently and significantly outperforms all baselines in terms of accuracy and F1.
PP-GCN achieves 18\%-37\% improvements in terms of accuracy and F1 in event classification tasks over all baselines, and 20\%-36\% improvements in terms of accuracy and F1 in event classification tasks over all baselines.

\begin{table}[t]
\caption{\label{tab:event-classification}Accuracy and F1 results of algorithms on the 2000 Weibo Data.}\center
\renewcommand{\multirowsetup}{\centering}
\begin{tabular}{|c|cc|cc|}
\hline
\multirow{2}{*}{Algorithms} & \multicolumn{2}{c}{Event} & \multicolumn{2}{c|}{Evolution}\\
\cline{2-5}
 & Accuracy & F1 & Accuracy & F1\\
\hline
SVM &  0.6477 & 0.6368 & 0.6654 & 0.6209\\
ARC-I~\cite{hu2014convolutional} & 0.6739 & 0.6693 & 0.6916 & 0.6455 \\
ARC-II~\cite{hu2014convolutional} & 0.7018 & 0.6802 & 0.7080 & 0.6538 \\
DUET~\cite{mitra2017learning} & 0.6684 & 0.7016 & 0.6753 & 0.7274 \\
MV-LSTM~\cite{wan2016deep} & 0.5574 & 0.6107 & 0.5985 & 0.6492 \\
C-DSSM~\cite{shen2014learning} & 0.6703	& 0.6956 & 0.6806 & 0.6936 \\
DSSM~\cite{huang2013learning} & 0.7039 & 0.7173 & 0.7218 & 0.7349 \\
SE-GCN~\cite{liu2018matching} & 0.7304 & 0.7518 & 0.7475 & 0.7519 \\
\hline
PP-EW-GCN & 0.7038 & 0.7171 & 0.7288 & 0.7371\\
PP-SE-GCN& 0.7621 & 0.7711 & 0.7737 & 0.7883\\
PA-GCN & 0.8668 & 0.8792 & 0.8891 & 0.8945\\
PP-GCN& \bf{0.9241} & \bf{0.9334} & \bf{0.9581} & \bf{0.9594}\\
\hline
\end{tabular}
\end{table}

\begin{figure*}[t]
\center
\includegraphics[width=0.9\textwidth]{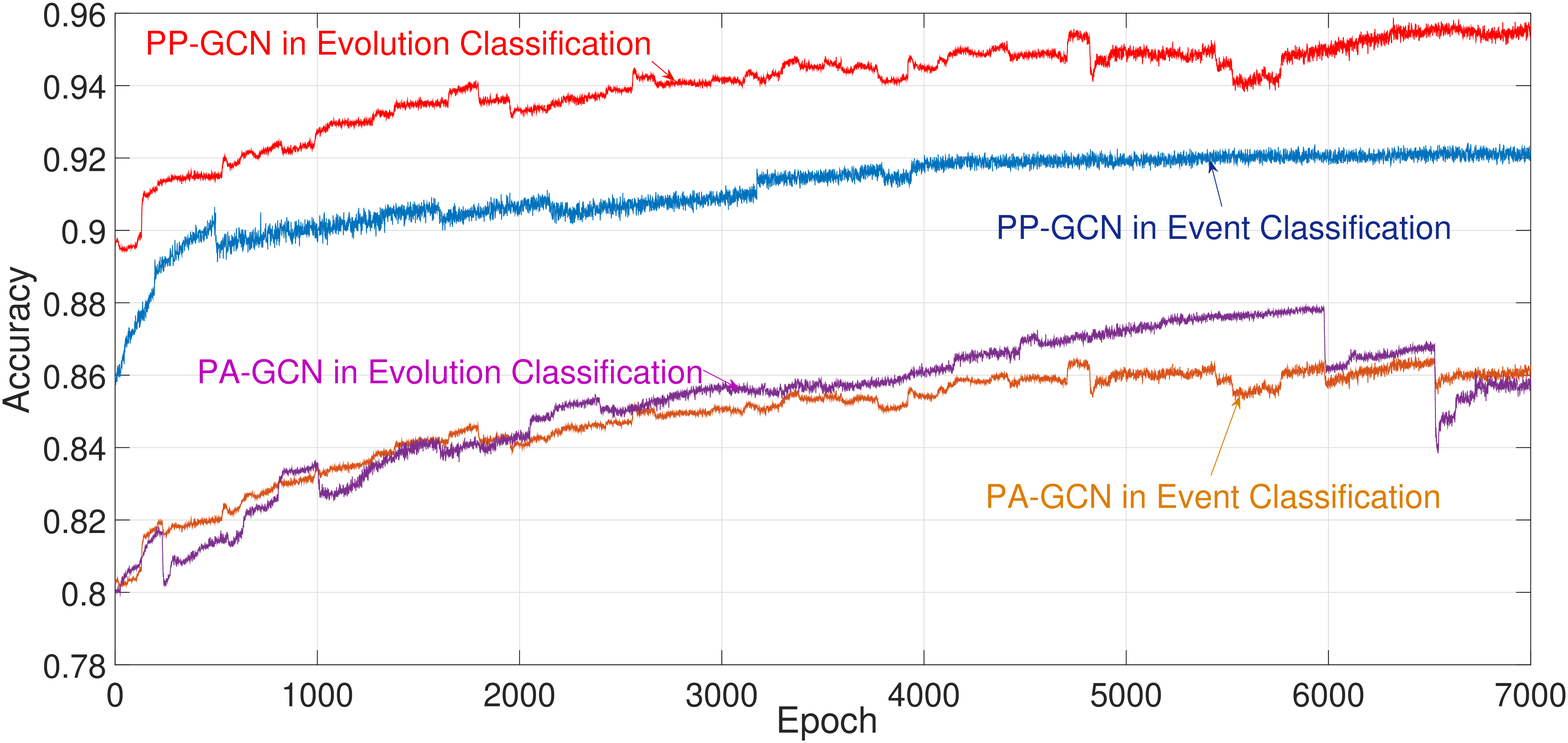}
\caption{Illustration of the Accuracy for PP-GCN and PA-GCN.}\label{fig:gcn_acc}
\end{figure*}

The improvements can be attributed to the three characteristics of proposed models.
First, the knowledgeable HIN is better modeling social events than traditional text modeling methods, such as bag-of-words (SVM), N-gram (ARC-I, ARC-II and C-DSSM) and sequence-of-words (MV-LSTM).
Our PP-GCN has improved overall by more than 18\% and 20\% in both event detection and evolution discovery classification tasks over the SE-GCN model incorporating structural and conceptual semantics.
Second, the combination of \emph{KIES} based weighted adjacency matrix and Doc2Vec is better for fine-grained event instance or event representation learning than for feature extraction on text pairs, such as DUET.
Third, the classifier based on the ratio of modulus of generated representations of event instances is better than the traditional pairwise distances.
Here, we replace the regression module of the SE-GCN model by our proposed popularity based classifier, named as PP-SE-GCN, and the performances can be improved 2\%-3\% in the Weibo data.
The 18\%-20\% improvements from the SE-GCN to the PP-GCN demonstrate the advantages of knowledgeable HIN modeling and the pairwise popularity based feature learning framework.
Furthermore, the proposed PP-GCN model can avoid overfitting in training.
We replace our classifier in PP-GCN by the discrete cosine angle of generated event instances that feature vectors based classifiers, namely PA-GCN.
In Figure~\ref{fig:gcn_acc}, we visualize the test accuracies of the PP-GAN and PA-GCN in both event classification and event evolution classification in 7000 epochs.
Compared to the angle-based classifier, the popularity-based classifier has better ability to learn discriminating and stable event instance features and prevent overfitting.

In order to verify that the learnable weights $\vec{\omega}$ are beneficial to social event representation, we test a PP-EW-GCN model with equal meta-path weight settings.
Compared with the models with learnable weights, the accuracy of the PP-EW-GCN method is significantly lower, as shown in Table~\ref{tab:event-classification}.
The reason is that the PP-EW-GCN model does not consider the importance of different meta-paths, that is, the importance of event elements such as different entities, keywords, topics, users, etc.
Compared with the SE-GCN model, the accuracy of the PP-EW-GCN model is reduced by 2\%-3\%. 
This comparative experiment also illustrates the importance of considering different elements of social events.

There are two advantages of the proposed PP-GCN model.
The one advantage is that we can perform highly accurate social event categorization.
The other advantage of the proposed PP-GCN compared to other methods is that the weights $\vec{\omega}$ between the meta-paths can be learned according to the event or evolution classification tasks.
So, we calculate the value of similarities between the same event class and the difference event classes, and evaluate the given values of the distance threshold $\epsilon$ in both event detection and event evolution tasks.

\begin{figure*}[t]
\centering
\includegraphics[width=0.9\textwidth]{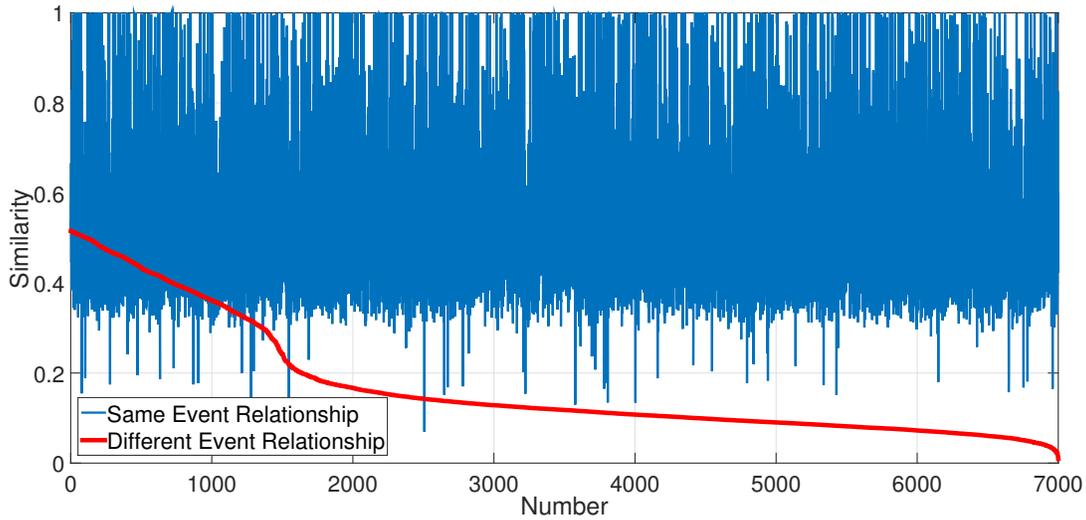}
\caption{Value of pairwise event similarity.}\label{fig:event_distribution}
\end{figure*}

\begin{figure*}[t]
\centering
\includegraphics[width=0.9\textwidth]{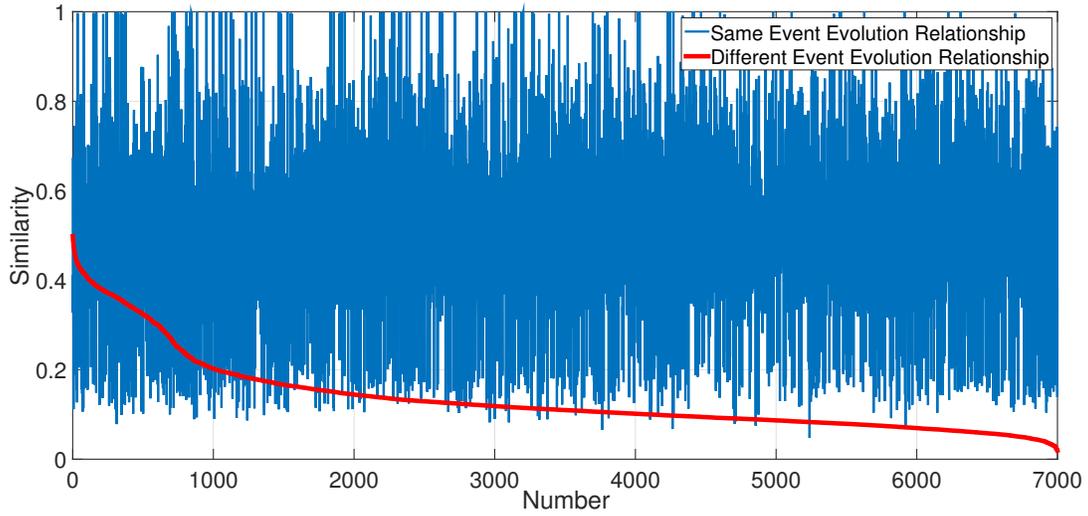}
\caption{Value of pairwise event evolution similarity.}\label{fig:evolution_distribution}
\end{figure*}

First, we randomly select 7000 social messages and 7000 social events, respectively.
Second, we randomly select both a sample belonging to the same event and a sample belonging to different events for each sample in the above 7000 social messages to build 14000 pairs of social messages.
Similarly, we randomly selected 2 pairs of samples for each of the 7000 social events to build 14000 pairs of social events.
Third, we calculate the similarity distance between pairs.
The values of pairwise event similarity and pairwise event evolution similarity are shown in Figure~\ref{fig:event_distribution} and Figure~\ref{fig:evolution_distribution}.
Here, we rank the similarities of different event or event evolution relationships from large to small.
The red lines represent the value of pairwise similarities in the different relationships.
The blue lines represent the value of the pairwise similarity of the same event or event evolution relationship.

\begin{figure*}[t]
\center
\includegraphics[width=0.9\textwidth]{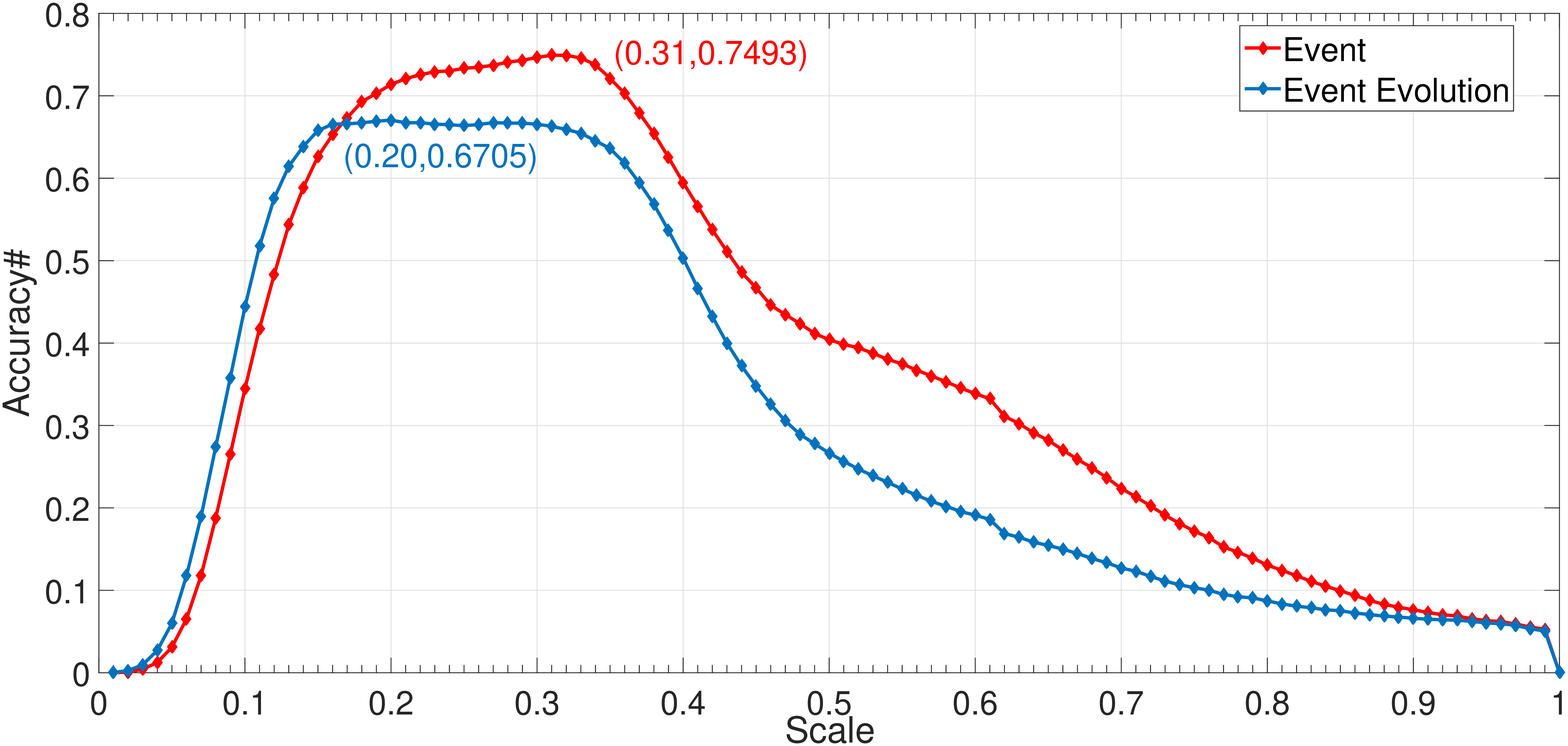}
\caption{Accuracy under different similarity thresholds.}\label{fig:thresholds_distribution}
\end{figure*}

In order to obtain best similarity thresholds for event and event evolution, respectively, we use an enumeration from 0 to 1 with a scale of 0.01 growth to evaluate the accuracies of event detection and evolution.
Here, if the similarity between two social events is greater than the selected threshold, then the two events are considered to be the relationship of the same event or the evolution of the same event.
The two accuracies under different thresholds are shown in Figure~\ref{fig:thresholds_distribution}.
We can see that the highest accuracy of event detection and event evolution can be achieved when the similarity thresholds are $0.31$ and $0.20$, respectively.
Due to the interpretability of the meta-path and similarity measure \emph{KIES}, the learned weights $\vec{\omega}$ can be utilized in the following event clustering based applications.
Therefore, we can use $0.69$ and $0.8$ as the given value of the $\epsilon$ in the next H-DBSCAN based event clustering.
We note that the above enumerating similarity is a simple pairwise distance test to obtain best $\epsilon$ thresholds for the H-DBSCAN based clustering experiments.

\subsection{Effectiveness and efficiency evaluation of static event detection and evolution discovery}
Next, we present the event clustering to evaluate the effectiveness and efficiency of event-based HIN framework and the \emph{KIES} distance measure in event detection and evolution discovery with the H-DBSCAN technology.
Due to the interpretability of the meta-path, we use the best weights $\vec{\omega}$ trained in Section~\ref{sec:weightslearning} for the calculation of \emph{KIES} in next experiments.
Another goal of this experiment is to find the best scales of the social messages and social events in one time slice in the streaming scenario.
Since the average daily acquisition of non-noisy and non-repetitive microblogs on distributed data collection platform is $2.35$ million, the average collection number in half an hour is about $48.9$ thousands of microblogs, and the microblogs data collected within half an hour ranges from a minimum of $10.4$ thousands to a maximum of $60.6$ thousands. 
We choose the one million manually labeled Sina microblogs dataset, and randomly choose 10000, 20000, 30000, 40000, 50000, 60000, 70000, 80000, 90000 and 100000 microblogs to construct different sizes of event-based HINs.

\begin{table*}[t]\caption{\label{tab:time_cost_hin_matrix}Time consumption for building different scales of event-based HINs and weighted adjacency matrixes (Minutes).}\center
\begin{tabular}{|c|cccccccccc|}
\hline
Items & 10,000 & 20,000 & 30,000 & 40,000 & 50,000 & 60,000 & 70,000 & 80,000 & 90,000 & 100,000\\
\hline
EE & 0.1610 & 0.3278 & 0.5082 & 0.7145 & 0.9391 & 1.1090 & 1.1793 & 1.3634 & 1.5526 & 1.7579 \\
ED & 0.8933 & 1.7416 & 2.5771 & 3.5851 & 4.6898 & 5.5140 & 6.0105 & 7.1530 & 7.259 & 8.0202\\
RK & 0.0438 & 0.0727 & 0.0996 & 0.1143 & 0.1356 & 0.1529 & 0.1703 & 0.1837 & 0.1991 & 0.2026\\
RE & 0.1442 & 0.1457 & 0.1468 & 0.1474 & 0.1485 & 0.1494 & 0.1501 & 0.1515 & 0.1520 & 0.1523\\
RKE & 0.2106 & 0.4289 & 0.6692 & 0.8035 & 1.0041 & 1.2456 & 1.4829 & 1.6999 & 1.9012 & 2.7690\\
Others  & 0.1520 & 0.2946 & 0.4351 & 0.5835 & 0.6976 & 0.7855 & 0.8483 & 0.9681 & 1.0938 & 1.3942 \\
\hline
Adjacency matrix & 0.1562 & 0.6250 & 1.4062 & 2.504 & 3.9062 & 5.6251 & 7.6562 & 10.0137 & 12.6562 & 15.6253 \\
\hline
Total time & 1.7613 & 3.6364 & 5.8421 & 8.4483 & 11.5209 & 14.5815 & 17.4978 & 21.5197 & 24.8139 & 29.2655\\
\hline
\end{tabular}
\end{table*}

We first build the event-based HIN, and we use one server with 64 processors and 1000 multi-threads to process social message texts.
Here, we use the dictionary structure to store all elements and relationships in the event-based HINs.
Second, in order to speed up the construction of the adjacency matrix, we divide the upper triangular matrix of the adjacency matrix into three triangular regions of the same size, and use three servers to simultaneously calculate the \emph{KIES} similarity between social message texts with the learned weights of meta-paths.
The detailed time consumption for constructing the different scale of event-based HINs and adjacency matrixes is shown in Table~\ref{tab:time_cost_hin_matrix}.
On the whole, the larger the social message texts, the more time it takes.
We note that the relationship between total time consumption and scale of social datasets approximates a trend of super-linear growth.
When constructing the event-based HIN, we report in detail the time consumption on 6 different phases, including entity extraction (EE), entity disambiguation (ED), relation between keywords (RK), relation between entities (RE), relation between keyword and entity (RKE) and others, where the most time consuming one is entity disambiguation.
Because social relationships, relationships between topics, and relationships between topic and words are extracted and pre-stored, we report the time consumption of identifying topics for social message texts, keyword extraction, etc, as others in Table~\ref{tab:time_cost_hin_matrix}.

\begin{figure*}[t]
\centering
\includegraphics[width=0.9\textwidth]{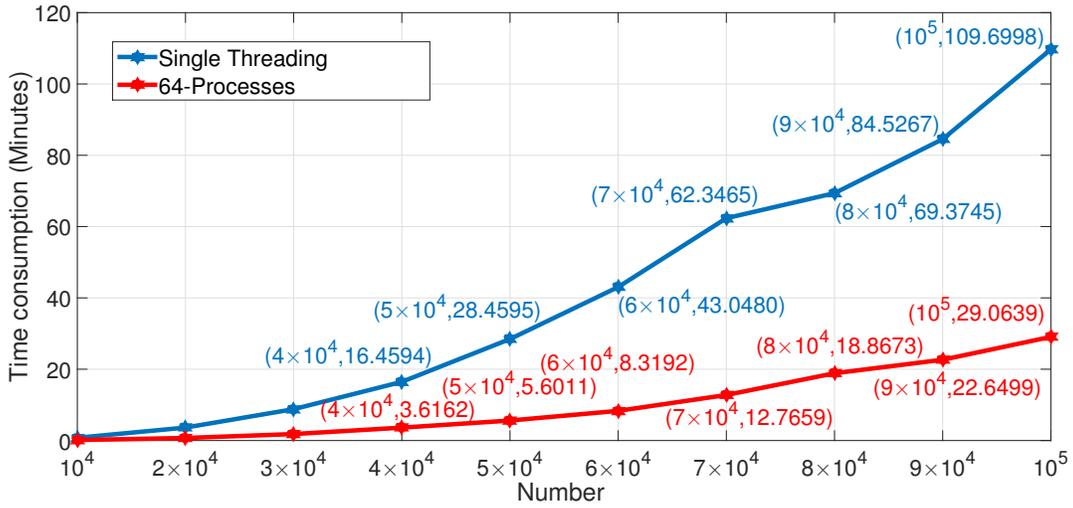}
\caption{Time consumption of event detection clustering for different sizes of social messages based on the parallel H-DBSCAN.}\label{fig:static_event_time}
\end{figure*}

\begin{figure*}[t]
\centering
\includegraphics[width=0.9\textwidth]{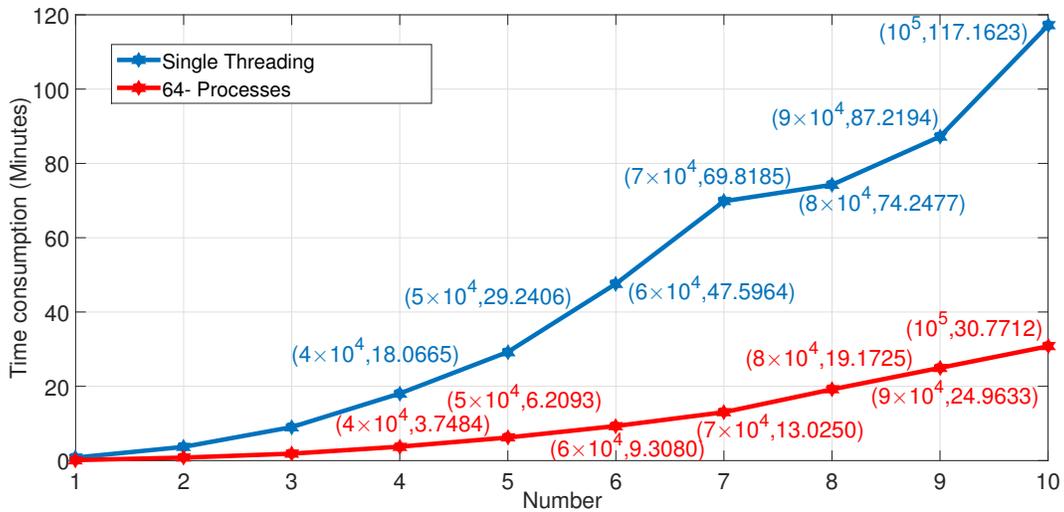}
\caption{Time consumption of event evolution clustering for different sizes of social events based on the parallel H-DBSCAN.}\label{fig:static_evolution_time}
\end{figure*}

After building the adjacency matrix, we use the parallel H-DBSCAN method, discussed in Section~\ref{sec:dbscan}, to perform event detection and evolution discovery, respectively.
Here, we use one server using maximum 64 processors and 1000 multi-threads to test the above parallel clustering experiments. 
The time consumption for event detection and evolution discovery are shown in Figure~\ref{fig:static_event_time} and Figure~\ref{fig:static_evolution_time}.
Blue lines represent the time-consuming using a single core, and red lines represent the time-consuming using 64 cores.
The speedup of 64 processors parallel H-DBSCAN ranges from 3.493 to 6.1184.
We mark the main time consumption and size of social messages in Figure~\ref{fig:static_event_time} and Figure~\ref{fig:static_evolution_time}.
As the average collection number in half an hour is about $48.9$ thousands of microblogs, we can see that the time consumption is about $11.5209$ minutes in building the corresponding event-based HIN and weighted adjacency matrix when the number of microblogs is $5\times 10^4$.
When we add all time consumption of the above processing and event detection clustering by the H-DBSCAN, $5\times 10^4$ social microblogs takes about $17.1219$ minutes.
Meanwhile, as the microblogs data collected within half an hour ranges from a minimum of $10.4$ thousands to a maximum of $60.6$ thousands, the time consumption ranges from about $1.3$ minutes to $8.3192$ minutes in the event detection clustering task.
When the number of social microblogs is $5\times 10^4$, the time consumption of the event evolution task takes $6.2093$ minutes.
When we add the time consumption of the above processing and evolution discovery by the H-DBSCAN, $5\times 10^4$ social microblogs takes about $17.7302$ minutes.
In addition to the efficiency of the proposed event-based HIN framework and H-DBSCAN, another important evaluation is the effectiveness of event detection and evolution discovery.

\begin{table}[t]\caption{\label{tab:event-clustering-dbscan}NMI results for event detection and evolution discovery on different number of social microblogs by the proposed H-DBSCAN.}\center
\renewcommand{\multirowsetup}{\centering}
\begin{tabular}{|c|cc|}
\hline
Items & Event Detection & Event Evolution\\
\hline
10,000 & 0.9365 & 0.9467 \\
20,000 & 0.9394 & 0.9402 \\
30,000 & 0.9432 & 0.9515 \\
40,000 & 0.9413 & 0.9411 \\
50,000 & 0.9337 & 0.9421 \\
60,000 & 0.9375 & 0.9345 \\
70,000 & 0.9479 & 0.9449 \\
80,000 & 0.9413 & 0.9440 \\
90,000 & 0.9344 & 0.9364 \\
100,000 & 0.9394 & 0.9355 \\
\hline
\end{tabular}
\end{table}

\begin{table}[t]\caption{\label{tab:event-clustering-metric}NMI results of different similarity metrics based DBSCAN for event detection and evolution discovery on the 50,000 microblogs.}
\center
\renewcommand{\multirowsetup}{\centering}
\begin{tabular}{|c|cc|}
\hline
Methods & Event Detection & Event Evolution\\
\hline
TF-IDF & 0.7002 & 0.6890 \\
LDA~\cite{blei2003latent} & 0.7195 & 0.7300 \\
mSDA~\cite{chenmarginalized} & 0.7658 & 0.7538 \\
CCG~\cite{perina2013documents} & 0.8172 & 0.8304\\
WMD~\cite{kusner2015word} & 0.8583 & 0.8326\\
KnowSim~\cite{wang2016text} & 0.8531 & 0.8324\\
\hline
KIES & \bf{0.9337} & \bf{0.9421} \\
\hline
\end{tabular}
\end{table}

\begin{table}[t]\caption{\label{tab:event-clustering-metric-kmeans}NMI results of different similarity metrics based K-means for event detection and evolution discovery on the 50,000 microblogs.}
\center
\renewcommand{\multirowsetup}{\centering}
\begin{tabular}{|c|cc|}
\hline
Methods & Event Detection & Event Evolution\\
\hline
TF-IDF & 0.6536 & 0.6445 \\
LDA~\cite{blei2003latent} & 0.6679 & 0.6488 \\
mSDA~\cite{chenmarginalized} & 0.7020 & 0.6837 \\
CCG~\cite{perina2013documents} & 0.7307 & 0.7071\\
WMD~\cite{kusner2015word} & 0.7866 & 0.7628\\
KnowSim~\cite{wang2016text} & 0.8079 & 0.7801\\
\hline
KIES & \bf{0.8412} & \bf{0.8211} \\
\hline
\end{tabular}
\end{table}

On the one hand, we evaluate the performances of NMI on event detection and event evolution tasks at different numbers of social microblogs from $10,000$ to $100,000$ by the proposed H-DBSCAN.
Here, we evaluate both event detection and event evolution tasks by calculating the purity of social microblogs belonging to the event and event evolution, respectively.
As shown in Table~\ref{tab:event-clustering-dbscan}, the proposed H-DBSCAN has stable performances in both event detection and event evolution tasks on different numbers of social microblogs.
For event detection clustering, the highest NMI is $0.9479$ when the number of social microblogs is $70,000$.
And for event detection clustering, the highest NMI is $0.9515$ when the number of social microblogs is $30,000$.
From these experiments, we can see that when the number of microblogs is larger, the more events and events evolution will be contained.
We also can conclude that the interaction between events is stable in our proposed framework and algorithms.
We also compare the proposed similarity measure \emph{KIES} with six popular document similarity metrics with the same DBSCAN based event detection and evolution discovery.
The experimental results are shown in Table~\ref{tab:event-clustering-metric}, our proposed similarity measure \emph{KIES} based DBSCAN achieves the best performances on the two clustering tasks in terms of NMI.
Moreover, among the baselines, the WMD, mSDA and CGG measures have been verified to achieve state-of-the-art effects in text similarity in ~\cite{kusner2015word}.
Although the KnowSim ignores the influence of social users and the weight given is rough, the KnowSim based DBSCAN also achieves performances of 85.31\% and 83.24\% in terms of NMI.
Compared to other similarity measures, the proposed \emph{KIES} based H-DBSCAN method achieves $8.06\%$-$25.31\%$ improvements in terms of NMI.
So, the proposed event-based HIN, \emph{KIES} and H-DBSCAN are highly scalable and effective.

In addition to DBSCAN based on density clustering, we also test the clustering detection effect of K-means under different similarity measures in Table~\ref{tab:event-clustering-metric-kmeans}.
We see that even if the same similarity measure is used, the clustering detection effect based on K-means is generally lower than that based on DBSCAN.
For example, even if the same Word Move Distance is used, the DBSCAN method consistently improves the accuracy of NMI by 7\% compared to the K-means in event detection and evolution discovery tasks.
Under our proposed \emph{KIES} measure, the DBSCAN method has about 9\% advantage over the K-means method in the accuracy of NMI.
This comparative experiment confirms that social event detection and evolution are more suitable for density clustering methods.
The same event or the evolution tends to be cohesive, and there is a discernible distance between the semantics of different events or evolutions.
In addition to accuracy, even in online social event detection and evolution discovery tasks, there is no pre-specified number of clusters.

\begin{table}[t]\caption{\label{tab:event-clustering}NMI results of different event detection and evolution discovery algorithms on the 50,000 microblogs.}\center
\renewcommand{\multirowsetup}{\centering}
\begin{tabular}{|c|cc|}
\hline
Methods & Event Detection & Event Evolution\\
\hline
HPA-ED~\cite{Marcus2011Twitinfo} & 0.4196 & - \\
CSCB-ED~\cite{Nguyen2017Real} & 0.5580 & - \\
SBAS-ED~\cite{yu2017ring} & 0.7121 & - \\
IIC-ED~\cite{hasan2019real} & 0.4218 & - \\
SIDF~\cite{weiler2014event} & 0.4079 & 0.5325 \\
eTrack~\cite{Lee2013Event} & 0.6408 & 0.6783 \\
SKCG~\cite{sayyadi2009event,sayyadi2013graph} & 0.6509 & 0.6917 \\
CSTP-EE~\cite{nallapati2004event} & - & 0.4551 \\
WLSH-EE~\cite{lu2015discovering} & - & 0.6935 \\
\hline
H-DBSCAN & \bf{0.9337} & \bf{0.9421}  \\
\hline
\end{tabular}
\end{table}

On the other hand, as the average $48.9$ thousands of microblogs in half an hour, we compare the performances of different event detection and evolution discovery algorithms on the basis of the $50,000$ social microblogs, as shown in Table~\ref{tab:event-clustering}.
Since the hashtag-based peak aggregation method only considers the hashtag information to cluster social events, the HPA-ED model can capture hot events, but the NMI result is relatively low.
Compared to the proposed H-DBSCAN method, the CSCB-DE model combines both content features and the propagation of social microblogs to detection events, but lacks modeling entities and relationships that are more semantically characterized.
The result of the CSCB-ED model is $0.5580$ in terms of NMI.
The SKCG method models social microblogs as a keywords co-occurrence graph-of-words, and then measures the similarity between microblogs by integrating both contextual and structural features to detect and track dense subgraphs.
The NMI results of SKCG method are $0.6509$ and $0.6917$ in terms of event detection and event evolution, respectively.
Similar to modeling social microblogs as a co-occurrence network, the SBAS-ED method builds the graph-of-words based on trending keywords, and detects the dense subgraphs as emergency events.
Compared to the proposed H-DBSCAN method, both SKCG and SBAS-ED methods are typical isomorphic anomaly subgraph detection models.
The IIC-ED model proposes a low computational cost or fast clustering method to detect social events. 
The result of the IIC-ED model is $0.4218$ in terms of NMI.
The SIDF model extracts the IDF feature for social microblogs, and uses a sliding window model to extract events.
The NMI results of the SIDF method are $0.4079$ and $0.5325$ in terms of event detection and event evolution tasks, respectively.
The eTrack method models social microblogs as evolving networks, where each social post is a node, and edges between posts are constructed when the post similarity is above a threshold.
The eTrack method clusters dense subgraphs as social events, and takes temporal evolving subgraphs as event evolution.
The NMI results of eTrack method are $0.6408$ and $0.6783$ in terms of event detection and event evolution tasks, respectively.
The CSTP-EE method proposes an event similarity that integrates both words and temporal locality of stories features to capture time-ordering dependencies among events.
The event evolution result of the CSTP-EE model is $0.4551$ in terms of NMI.
The WLSH-EE method evaluates the evolutionary relation between events by modeling a weighted linear similarity function based on defined event elements.
The event evolution result of the WLSH-EE model is $0.6935$ in terms of NMI.

Overall, the proposed event-HIN framework, \emph{KIES} and H-DBSCAN algorithm consistently and significantly outperforms all baselines in terms of NMI in both event detection and event evolution tasks. 
In the $50,000$ social microblogs, the NMI results of H-DBSCAN achieve $52.58\%$ and $48.70\%$ improvements in event detection and event evolution, respectively.
The above comparative experiments and time cost consumption analysis demonstrate both effectiveness and efficiency of the proposed framework and model.
Next, we evaluate the performances of streaming social event detection and evolution discovery with the proposed framework and model.

\subsection{Effective and efficiency evaluation of streaming event detection and evolution discovery}
Different from static social event mining, streaming social event detection and evolution clustering should firstly solve the problem of event merging.
For example, the tweet associated with a social event will continue to spread over the social network for a period of time as the event heats up.
Therefore, we retrieve relevant social message texts from neighboring time for current social message texts.

We have modeled social messages and events as the event-based HIN, as shown in Figure~\ref{fig:event-hin}.
Therefore, we can retrieve relevant social message text on the HIN through simple meta-paths, as shown in Figure~\ref{fig:metapaths}.
Here, we also note that there are multiple instances of the above two meta-paths.
We use 64 cores to retrieve relevant social message texts of different sizes ranging from $10,000$ to $50,000$, and employ the meta-paths enumerated from IEI and EIMIE, respectively, for event detection and event evolution tasks.
We remove duplicate social microblogs.
The time consumption for the meta-paths based relevant messages or events searching is shown in Table~\ref{tab:time_cost_hin_retrieve}.
We can see that time consumption is acceptable and relatively short.
When we retrieve $50,000$ unique social messages or events, it takes $1.7712$ or $1.9690$ minutes.

\begin{table}[h]\caption{\label{tab:time_cost_hin_retrieve}Time consumption for different scales of social messages searching on the event-based HIN (Minutes).}\center
\begin{tabular}{|c|ccccc|}
\hline
Items & 10,000 & 20,000 & 30,000 & 40,000 & 50,000\\
\hline
Metapath:  IEI & 0.3052 & 0.635 & 0.989 & 1.3253 & 1.7712 \\
Metapath: EIMIE & 0.3415 & 0.694 & 1.1755 & 1.5875 & 1.9690 \\
\hline
\end{tabular}
\end{table}

\begin{figure*}[t]
\centering
\includegraphics[width=0.9\textwidth]{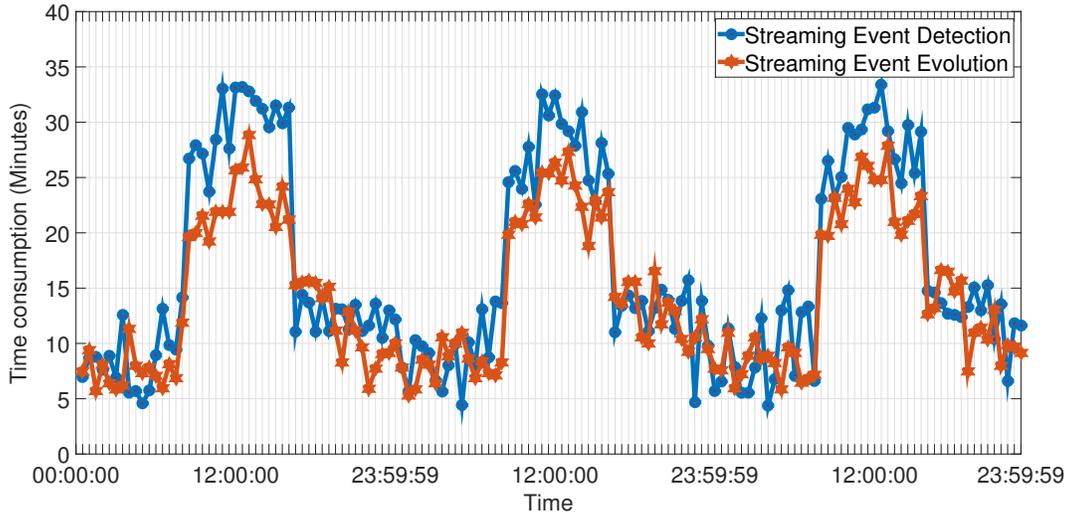}
\caption{The time consumption of streaming event detection and event evolution within 3 days.}\label{fig:stream_time}
\end{figure*}

In the streaming scenario, we also construct the streaming event-HIN while collecting social messages, and perform the H-DBSCAN based event detection and event evolution for half an hour, as shown in Figure~\ref{fig:architecture}.
Although the average collection number in half an hour is about $48.9$ thousands of non-noisy and non-repetitive microblogs, and the microblogs data collected within half an hour ranges from a minimum of $10.4$ thousands to a maximum of $60.6$ thousands.
We also retrieve the similar microblogs on the neighboring time slice for streaming microblogs to build the temporal event-based HIN, where more than a week of information is stored in the external storage space.
Then, we performed H-DBSCAN based event detection and evolution discovery, respectively, on two servers.
The time consumption of streaming event detection and event evolution within 3 days are shown in Figure~\ref{fig:stream_time}.
We find that the number of microblogs collected and retrieved at each time changes over time.
In the overall trend, we observe that the number of postings by social users during the day is relatively large, and the number of postings at night is small.
For streaming event detection, the maximum time consumption for each time slice is about $33.15$ minutes. 
For streaming event evolution, the maximum time consumption for each time slice is about $28.76$ minutes. 
Different from the static event detection and event evolution, since the number of events is generally less than microblogs in one time slice, and part of event-based HIN has built in the event detection, the average of time consumption of streaming event evolution is less than the average of time consumption of streaming event detection.
Overall, our proposed streaming event-based framework and H-DBSCAN algorithm can meet real-time requirements.

\begin{figure}[t]
\centering
\includegraphics[width=0.9\textwidth]{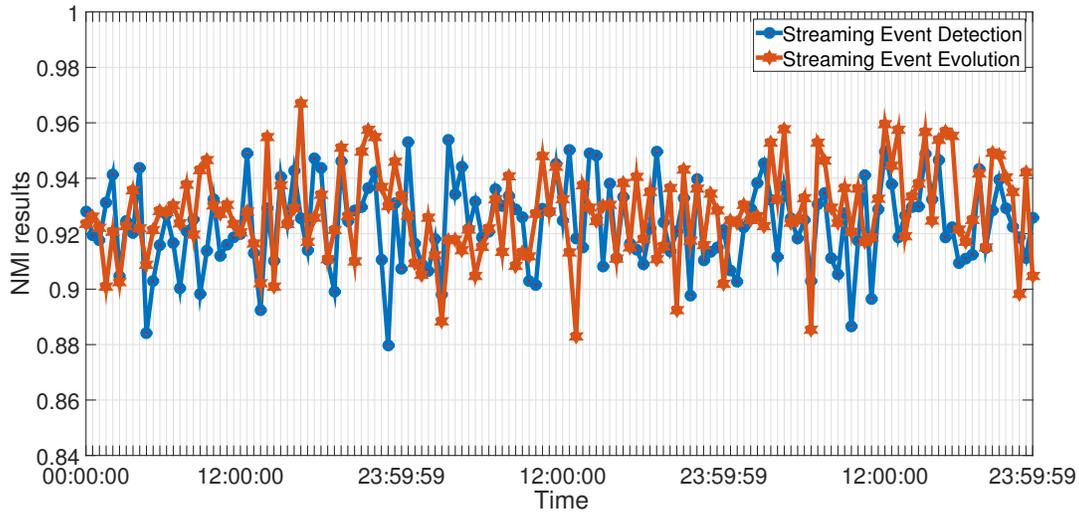}
\caption{The NMI results of streaming Event detection and event evolution within 3 days.}\label{fig:stream_nmi}
\end{figure}

In addition to the real-time requirements of streaming computing, we also mix a random number of manually annotated social microblogs into the streaming social data to measure the NMI of streaming event detection and evolution discovery.
The results of the event detection and event evolution are shown in Figure~\ref{fig:stream_nmi}.
In the streaming experiments, the performances are fluctuating, but the average NMI results are about $0.9233$ and $0.9275$.
For the streaming event detection task, the highest NMI is $0.9539$, and the lowest NMI is $0.8796$.
Meanwhile, for the streaming event evolution task, the highest NMI is $0.9670$, and the lowest NMI is $0.8829$.
The reason is that the incompleteness of the mixed annotation manually annotated social microblogs affects the number of points required to form a cluster in the H-DBSCAN.
Overall, the performances of event detection and evolution discovery are satisfactory for building streaming event detection and evolution systems.

\begin{table*}[t]\caption{Case Study.}\label{tab:case}
\begin{center}\small
\begin{tabular}{p{0.40\columnwidth}p{0.32\columnwidth}p{0.3\columnwidth}}
\toprule
Related Social Messages & Event Detection & Event Evolution\\  
\midrule
2019/04/15 Watch: Notre Dame Cathedral in Paris on fire. $\cdots$
& Event-1: \underline{2019/04/15 18:50$\sim$2019/04/16 09:36} Paris' Notre Dame Cathedral on fire.
& \\
2019/04/15 Notre Dame fire: Live updates.
& Time : 2019/04/15 18:50
&
\\
2019/04/15 400 firefighters mobilized for Notre Dame blaze.
& Location : Notre Dame Cathedral, Paris, French
&
\\

2019/04/15 Around the world, 'our hearts ache' at Notre Dame Cathedral fire.
& Topics : Fire disaster, Culture, Religion
& \underline{2019/04/15 18:50$\sim$2019/04/16 09:36} Paris' Notre Dame Cathedral on fire.
\\
2019/04/16 The church is burning and the whole world is crying - Parisians mourn for Notre-Dame.
& Entities : Notre Dame Cathedral, French President, Paris Fire Department, Rose Windows, Ministry of Culture, Pipe Organs $\cdots$
&
\\
$\cdots$\\
2019/04/15 Notre Dame cathedral fire is fully extinguished.
& Event-2: \underline{2019/04/15 23:15$\sim$2019/04/21 11:35} Notre Dame fire is extinguished.
& \\
2019/04/16 Why the Notre Dame fire was so hard to put out.
& Time : 2019/04/15 23:15
&
\\
2019/04/16 The fire is out': Paris firefighters succeed after 12-hour battle to extinguish Notre Dame Cathedral blaze.
& Keywords : Fire, Paris, Notre-Dame, Firefighter, Flame, Water, Helicopter, Tower, Heat, Stair, Edifice, Risk, Brigade, Police 
& \underline{2019/04/15 23:15$\sim$2019/04/21 11:35} Notre Dame fire is extinguished.
\\
2019/04/17 Paris firefighters got on Notre-Dame site in less than 10 minutes.
& Entities : Notre Dame Cathedral, Deluge Guns, Thermal Imaging, Bell Tower $\cdots$
&
\\
2019/04/21 Paris Easter Mass honors firefighters who saved Notre Dame.
& Topics : Fire disaster, Culture, Religion
&
\\
$\cdots$\\
2019/04/16 800 million euros in donations for the reconstruction of Notre-Dame. 
&Event-3: \underline{2019/04/16 07:45 $\sim$ 2019/04/18 09:23} Reconstruction of Notre Dame de Paris.
&  \\
2019/04/16 Macron said reconstruction of the Cathedral.
& Time : 2019/04/16 07:45
&
\\
2019/04/16 Companies and large fortunes mobilize for the reconstruction of Notre-Dame.
& Entities : Fondation du Patrimoine, Palace of Westminster, 2024 Olympic Games, French prime minister, Arnault Family, Total SA, 
& \underline{2019/04/16 07:45 $\sim$ 2019/04/18 09:23} Reconstruction of Notre Dame de Paris.
\\
2019/04/17 Pledges Reach Almost 1 Billion To Rebuild Paris' Notre Dame Cathedral.
& BPCE, Bettencourt Family, Pinault Family, AXA SA, Paris City Government $\cdots$
&
\\
2019/04/18 Opinion: The reconstruction of Notre-Dame is not the only answer.
& Keywords : rebuild, organisation, insurance, millions, euros, cathedral, architecture $\cdots$
&
\\
$\cdots$\\
\bottomrule
\end{tabular}
\end{center}
\end{table*}

With the effectiveness and efficiency, the proposed event-based HIN, \emph{KIES}, PP-GCN and H-DBSCAN solve the problems of building reliable and open domain streaming event detection and event evolution systems in practice. 
Here, we give a case study of the events of ``Notre Dame de Paris''.
From the real-world social microblogs, we give the related social messages, where we describe it with significant parts and represent some unimportant parts as ellipses.
As happened in the real-time, from social messages, we detect that the entire case consisted of three phases of events, \underline{2019/04/15 18:50$\sim$2019/04/16 09:36} Paris' Notre Dame Cathedral on fire, \underline{2019/04/15 23:15$\sim$2019/04/21 11:35} Notre Dame fire is extinguished, and \underline{2019/04/16 07:45 $\sim$ 2019/04/18 09:23} Reconstruction of Notre Dame de Paris.
So, we give parts of event elements in these events.
Both main participants and concerned entities are different in different events throughout the case.
However, all three events are caused by the event of \emph{Paris' Notre Dame Cathedral on fire}.
Obviously, there is an evolutionary relationship between the three events.
More cases can be viewed at our developed event detection and evolution discovery system\footnote{http://ring.act.buaa.edu.cn}.

\section{Related Works}\label{sec:related}
In this section, we will briefly review the related work about social events and HIN.
Social event detection and evolution models can be roughly categorized into topic detection and tracking models, social media event frames extraction, abnormality detection based event detection.

Topic detection and tracking (TDT) originally described the detection of topics in news streams~\cite{Allan:2012:TDT:2481012}.
At that time, the detected topics were termed ``events'' and detecting and tracking them was considered a classification or clustering problem. 
Bag-of-words (BoW) representation is commonly used to represent a news article and do machine learning in the vector space defined by the BoW's features. 
Later on, probabilistic latent semantic analysis (PLSA)~\cite{hofmann1999probabilistic}, latent Dirichlet allocation (LDA)~\cite{blei2003latent}, and other techniques came to be used. 
The assumption that one news article contains only one topic was relaxed and that improved topic detection results. 
Overall, these early approaches mostly assumed that the features of the news or events were just words. 
They did not consider entities outside the article and their relationships.
Unlike TDT, social event extraction~\cite{Agarwal2010,Liu2012,mao2021event} uses frame-based event definitions applying the well-defined techniques for extracting event frames from news in natural language processing. 
Frame-based event extraction can extract entities and their relationships, but uses only a limited number of event types. 
Moreover, it uses complicated machine learning models, usually a pipeline of them, to incorporate different levels of annotation and features. 
However, social messages are invariably short and usually noisy. 
That calls for more flexibility to incorporate simpler but more reliable features and links.
Sometimes event detection is treated as abnormal detection~\cite{chandola2009anomaly} or document clustering~\cite{liu2020story} in streaming data. 
Streaming data can be of different types. 
Here we will only review some graph-based problems. 
Streaming's abnormally connected subgraph structure (under different names such as k-clique, motifs or graphlets) based social event detection has been studied~\cite{aggarwal2012event,angel2012dense,yu2017ring}. 
A sharp rise in keyword occurrences~\cite{yu2013anomalous,liu2018event,liu2020event} in a specified time slice is an important characteristic of events shown up in social media, which generates statistically-significant subgraphs or patterns. 
Subgraph analysis can potentially help to partition graphs better in graph-of-words document representation~\cite{rousseau2015text} when not only considering hot words in the stream but also word co-occurrence patterns.
These approaches have been applied to homogeneous graphs but not HINs. \\
\indent A HIN is a graph of entities with multiple typed entities and relations~\cite{sun2012mining,he2019hetespaceywalk}. 
They were originally developed to analyze scientific publication networks~\cite{sun2011pathsim}, and in social network analysis they can also represent user similarity and predict links~\cite{Zhang2013}. 
Similar to our work, Wang et al~\cite{wang2018unsupervised} integrated information from external knowledge base to help modeling rich textual HIN, and showed that it can be very useful for categorizing documents. 
HIN can also be used in time-evolving clustering tasks~\cite{sun2014}. 
There have been some studies of using HIN to analyze events. 
Gui et al~\cite{Gui2017Large} focused on a particular type of action (publishing a paper) as an event.
Thus using HIN to fully model events' elements (such as keywords, entities, hashtags, location, time) and heterogeneous social networks (with different types of nodes and different kinds of social relations) in event mining problems has not yet received much scholarly attention.

\section{Conclusion}\label{sec:conclu}
We study the problem of streaming social event detection and evolution discovery in the microblogs streams.
We first propose an event-based HIN framework which integrates event elements and their relations, in a semantically meaningful way, and calculates the similarity between any two events to model social messages.
Secondly, we propose the PP-GCN model that achieves state-of-the-art results in fine-grained event categorization, and learns the best weights of meta-paths in different tasks.
Through the PP-GCN model, we are able to overcome the problems of large category size and sparse small number of samples per class and prevent overfitting in our tasks.
Third, we propose the H-DBSCAN clustering method which achieves state-of-the-art results in both static and streaming event detection and evolution discovery tasks.
Our extensive experiments have demonstrated that the event-based HIN framework and H-DBSCAN clustering are highly effective and efficient.
In the future, it is interesting to extend the streaming event detection and evolution framework for handling multilingual social streams and predicting future social events and its influences.

\section*{Acknowledgment}
The authors of this paper were supported by the NSFC through grants 62002007 and U20B2053, 
the Key Research and Development Project of Hebei Province through grant 20310101D,
NSF of Guangdong Province through grant 2017A030313339,
Hong Kong RGC including Early Career Scheme (ECS, No. 26206717), General Research Fund (GRF, No. 16211520) and Research Impact Fund (RIF, No. R6020-19),
State Key Laboratory of Software Development Environment (SKLSDE-2020ZX-12), 
the UK EPSRC (EP/T01461X/1, EP/T021985/1, EP/R033293/1 and EP/T022582/1),
NSF ONR N00014-18-1-2009,
NSF under grants III-1763325, III-1909323, and SaTC-1930941.
This work was also sponsored by CAAI-Huawei MindSpore Open Fund. 
Thanks for computing infrastructure provided by Huawei MindSpore platform.
\bibliography{sigproc}

\end{document}